\documentclass[12pt,preprint]{aastex}
\usepackage{times}
\newcommand{\etal}{{\em et al.}}            

\shorttitle{Radio through X-ray of 3C 33}
\shortauthors{Kraft \etal}

\begin{document}

\title{A Radio through X-ray Study of the Hot Spots, Active Nucleus, and Environment 
of the Nearby FR II Radio Galaxy 3C 33}
\author{R. P. Kraft}
\affil{Harvard/Smithsonian Center for Astrophysics, 60 Garden St., MS-67, Cambridge, MA 02138}
\author{M. Birkinshaw}
\affil{University of Bristol, Department of Physics, Tyndall Avenue, Bristol BS8 ITL, UK}
\author{M. J. Hardcastle}
\affil{University of Hertfordshire, School of Physics, Astronomy, and Mathematics, Hatfield AL10 9AB, UK}
\author{D. A. Evans}
\affil{Harvard/Smithsonian Center for Astrophysics, 60 Garden St., MS-67, Cambridge, MA 02138}
\author{J. H. Croston}
\affil{University of Hertfordshire, School of Physics, Astronomy, and Mathematics, Hatfield AL10 9AB, UK}
\author{D. M. Worrall}
\affil{University of Bristol, Department of Physics, Tyndall Avenue, Bristol BS8 ITL, UK}
\author{S. S. Murray}
\affil{Harvard/Smithsonian Center for Astrophysics, 60 Garden St., MS-2, Cambridge, MA 02138}

\begin{abstract}

We present results from {\em Chandra}/ACIS-S, {\em Spitzer}, {\em XMM-Newton}, 
{\em HST}, and VLA observations of the
radio hot spots, extended environment, and nucleus of the nearby ($z$=0.0597) FR II radio galaxy 3C 33.
This is a relatively low-power FR II radio galaxy, and
so we expect, {\it a priori}, to detect a significant X-ray
synchrotron component to the emission from the hot spots.
We detect X-ray emission coincident with the two knots of polarized optical emission
from the southern hot spot (SHS), as well as along the northwest arm of this hot spot.
We also detect X-ray emission from two compact regions of the northern hot spot (NHS),
as well as diffuse emission behind the radio peak.
The X-ray flux density of the region at the tip of the southern hot spot, the
most compact radio feature of the southern lobe, is consistent with the 
synchrotron self-Compton (SSC) process.
The X-ray flux densities of the other three regions of the SHS
and the two compact regions of the NHS
are an order of magnitude or more above the predictions from either
the SSC and inverse-Compton scattering of the CMB (IC/CMB) mechanisms,
thus strongly disfavoring these scenarios unless 
they are far from equipartition ($B$ $\sim$4-14 times smaller than the
equipartition values).
The X-ray flux from the diffuse region behind the NHS is consistent with the
IC/CMB prediction assuming a small departure from equipartition.
However, the radio to X-ray flux density distributions of the three regions of the
SHS cannot be fitted with a fully convex, monotonically
decreasing spectral model as a function of frequency.  
We conclude that the X-ray emission is synchrotron emission 
from multiple populations of ultrarelativistic electrons unless these
regions are far from equipartition.  There must therefore be complex, 
unresolved substructure within each knot, similar to that which is found in Chandra 
observations of nearby FR I jets such as Centaurus A and 3C 66B to explain the
flux density distributions.
The nuclear spectrum consists of a heavily absorbed ($N_H\!\sim$4$\times$10$^{23}$ cm$^{-2}$)
power law with a bright Fe K$_\alpha$ line ($EW\!\sim$200 eV)
and an Fe K edge, thermal emission from galactic
gas, and an additional component that could either be reflection from the torus or
a second power law representing emission from an unresolved pc scale jet.
The detection of X-ray emission from both hot spots combined with the
large absorbing column toward the primary power-law component of the nucleus
conclusively demonstrate that the jets must lie relatively close to the plane of 
the sky and that relativistic beaming cannot be important.
We also detect diffuse X-ray emission, some of which is likely to originate in hot
galaxy or group-scale gas, and some of which may be attributed to IC/CMB
off the relativistic electrons of the radio lobes.

\end{abstract}

\maketitle
\section{Introduction}

The origin of X-ray emission from the radio hot spots of FR II radio
galaxies is still a subject of considerable debate.  For many sources,
inverse-Compton scattering of radio photons by the relativistic electrons
of the hot spot (the synchrotron self-Compton mechanism (SSC)) with the
magnetic field at or near the equipartition value provides an adequate
description of the broad band (radio to X-ray) spectra \citep{har94,mjh02}.  
There are, however, many sources for which the X-ray emission is orders of magnitude
larger than the SSC prediction and for which this paradigm is clearly
incomplete.  It has recently been suggested, based on analysis of archival
{\em Chandra} observations of the compact hot spots of
3CRR radio galaxies, that in some cases synchrotron emission
can be the dominant component and can be orders of magnitude larger
than the SSC component \citep{mjh04}.
It was found that lower radio power hot spots tended to have stronger
optical and X-ray synchrotron components \citep{bru03,mjh04}.  It was
speculated that this represents an intermediate case between X-ray synchrotron 
emission as seen in the jets of FR I radio
galaxies and the SSC mechanism of the powerful FR II hot spots. 
Some hot spots do indeed appear to show a SED consistent with synchrotron
emission from a simple continuous-injection model \citep{hm87}
between the radio and X-ray (e.g. 3C\,403; \citealt{khwm05}).
However, \cite{tav05}~showed that good optical/UV constraints on the
fluxes of the hot spots in some FR II quasars preclude fitting a single-component
synchrotron model to the radio through X-ray data, while the
normalization of the X-ray emission is inconsistent (orders of
magnitude too large) with an SSC model.  Instead, they
suggested that models involving beaming (such as the deceleration
model of \citealt{gk03}) may be required to explain the observations.
The synchrotron hypothesis can be directly evaluated with high resolution
X-ray images of the hot spots of {\it nearby} FR IIs: high spatial
resolution is necessary if we are to understand the relationship
between the X-ray emission and the various radio and optical structures related to jet
termination in these sources.

In the paper, we present results from new {\em Chandra}, {\em Spitzer}, and {\em XMM-Newton}
observations, combined with archival {\em HST}, {\em GALEX}, and VLA observations,
of the hot spots, the  nucleus, and the extended environment
of the nearby ($z$=0.0597, see \citep{pop92}) FR II radio galaxy 3C 33.  
3C 33 is a double radio source that extends $\sim$4$'$ on the sky with a weak radio core \citep{har75}.  
The radio emission at 1.4 GHz is dominated by the southern hot spot (SHS) \citep{rud88,rud90}.
Optical emission coincident with the radio peak of the SHS was first reported by \citet{sim79},
and it was subsequently determined that this emission
is highly polarized ($\sim$40\%), thus conclusively
confirming it as synchrotron emission \citep{mei86,cra92}.
Imaging and spectroscopic studies of the hot spot have shown a complex optical morphology
with a steep power-law slope \citep{sim86,cra87}.
The hot spots were not detected in a ROSAT/HRI observation, but the upper limit
to the X-ray flux demonstrated that there must be a break in the synchrotron
spectrum between the optical and X-ray bands \citep{mjh98}.
The primary scientific goal of the new observations is to resolve spatially the emission regions
of the hot spots from the radio to the X-ray band in order to elucidate the emission processes
that are important at the terminal hot spots of FR IIs.  3C 33 is an ideal
candidate for this study as it is the nearest narrow-line FR II in the 3CRR catalog
with optical synchrotron emission detected from a hot spot \citep{mei86}.
The proximity of 3C 33 permits us to resolve spatially the emission features 
at the highest possible linear resolution.  In addition, since this source has a narrow-line 
nucleus, relativistic beaming is not likely to be important in modifying the
spectral or morphological properties of the hot spots.  

This paper is organized as follows.  Section 2 contains a summary of the
observations, the results of the data analysis are presented in section 3,
and we discuss the implications of our results in section 4.
Section 5 contains a brief summary and conclusions.
We assume WMAP Year-1 cosmology throughout this paper \citep{spe03}.  The observed redshift
($z$=0.0597) of the host galaxy of 3C 33
corresponds to a luminosity distance of 263.9 Mpc, and one arcsecond
is 1.14 kpc.  All uncertainties are at 90\%
confidence for one parameter of interest
unless otherwise stated, and all coordinates are J2000.
Absorption by gas in our Galaxy ($N_H$=4.0$\times$10$^{20}$ cm$^{-2}$) \citep{dic90} 
is included in all spectral fits and count rate to flux density conversions.

\section{Observations}

The radio galaxy 3C 33 was observed with {\em Chandra}/ACIS-S in two observations
(OBSID 6190 and 7200) on November 8, 2005 and November 12, 2005 (PI: Stephen Murray).
The total observing time was $\sim$40.3 ks.  The observation was pointed so that the SHS
was near the best focus, but all regions of the radio galaxy, including
the northern hot spot (NHS) and nucleus, were contained on the S3 chip.
We made light curves for the S3 CCD in the 5.0 to 10.0 keV band, excluding
the nucleus, to search for background flares, and removed intervals where the background rate
was more than 3$\sigma$ above the mean, leaving 39831.6 s of good data.
Bad pixels, hot columns, and events along node boundaries also were removed,
and standard ASCA grade filtering (0,2,3,4,6) was applied to the events file.
Emission from 3C 33 fills only a small fraction of the field of view of
the S3 chip, so local background was used in all spectral analysis.
The X-ray core is aligned to $\sim$0.1$''$ with the radio core, so
no adjustment of the X-ray coordinates was necessary.
All X-ray images presented in this paper were generated from the
{\em Chandra} data.

{\em XMM-Newton} observed 3C 33 for a total of 17 ks in two separate
observations on 2004 January 4 and 2004 January 21 (PI: Martin
Hardcastle). The data were
processed with the Scientific Analysis Software (SAS) using the
standard pipeline tasks {\it emchain} and {\it epchain} and filtered
for patterns $<= 4$ (MOS) or $<=12$ (pn) and using the bit-mask flags
0x766a0600 for MOS and 0xfa000c for pn, which are equivalent to the
standard flagset \#XMMEA\_EM/EP but include out of field-of-view
events and exclude bad columns and rows. In the first of the two
observations, the background count rate was low ($\sim$0.2 counts
s$^{-1}$ for MOS and $\sim$0.4 counts s$^{-1}$ for pn) so that no
temporal filtering was required; however, both the MOS and pn
data from the second observation suffered from a large flare at the
start of the observation, which rendered the pn data unusable and
required the MOS data to be filtered using a count rate threshold of
0.5 counts s$^{-1}$. The resulting clean, merged data had a duration
of 9700 s (MOS) and 6314 s (pn).
The {\em XMM-Newton} data were used primarily in our study of the nucleus
and the extended, low surface brightness diffuse emission from the group.
The SHS was detected in these observations, the NHS was not.  The larger
{\em XMM-Newton} PSF (relative to {\em Chandra}) provides no additional structural
constraints, so we use only the {\em Chandra} data in our study of the hot spots.

We present data from three {\em Spitzer} observations of 3C~33, two data sets
taken from the public archive (one IRAC and one MIPS observation, both
pointed towards the SHS - program ID 3327), and an additional
IRAC observation (program ID 3418) centered on the
host galaxy taken by the 3CRR low-z consortium (Birkinshaw \etal~2007,
in preparation). The IRAC observations were made on July 23, 2004 (3327), and
January 6, 2005 (3418), with observations times of 96 and 360 seconds,
respectively.  The MIPS (24 $\mu$m only) observation was made on
December 23, 2004 (3327) with an observation time of 200 seconds. 
The images used were produced by the data
analysis pipeline version 14.1.0.  The SHS is detected in all
four IRAC bands and the MIPS image. The NHS is only contained within 
the {\em Spitzer} field of view
in the 2005 IRAC observation and the MIPS image, and is detected
in all four IRAC bands and the MIPS 24 $\mu$m band, although there is some
confusion from adjacent stars.

We also use archival {\em HST} (WFPC and WFPC2) and VLA data in this paper.  
The SHS was observed for 2400 s with the {\em HST}/WFPC2 instrument
in 1995 as part of the {\em HST} survey of radio-galaxy hot spots
(PI: P. Crane) using the F702W filter (pivot wavelength of 6919 \AA), and
for 1800 s with the {\em HST}/WFPC instrument
using the F606W filter (pivot wavelength of 5888 \AA) \citep{cra92}.
There have been no {\em HST} observations of the NHS.
We obtained the reprocessed data of the SHS from the {\em HST} archive and used
the IRAF {\it synphot} package to apply photometric calibrations.
The fluxes were reddening-corrected using the dust maps of \citet{sch98}.
The correction to the flux densities is $\sim$25\% in the visual.
A 1.5-GHz radio map with a resolution of 4.0 arcsec was obtained from the 3CRR
Atlas\footnote{http://www.jb.man.ac.uk/atlas/}: this is the image of
\citet{lp91}. At higher frequencies, we used 4.9 and 15-GHz data from
the VLA public archive.  Details of the VLA data used and the maps
made from them are given in Table~\ref{vla}. The data were calibrated
and reduced in the standard manner using AIPS. Results from analyses
of these data have been previously published in \citet{rud88} and
\citet{rud90}.  

Finally, we also examined the {\em XMM-Newton} Optical
Monitor (UVM2 filter) data and archival {\em GALEX} 
images of 3C 33.  The hot spots were not detected in any of
the UV images.  The observation time of the {\em GALEX} observation
was sufficiently short ($\sim$90 s) that the upper limits are not interesting
(they lie well above an extrapolation of the optical detections
or limits).  The {\em XMM-Newton} OM upper limits lie above the {\em GALEX} 
limits because of a combination of the narrower bandpass and lower effective area.
The 3$\sigma$ upper limits on the flux density of S1
in the {\em GALEX} NUV (pivot wavelength 2297 \AA) and FUV 
(pivot wavelength 1524 \AA) bands are 2.0 and 6.1 $\mu$Jy, respectively.
These values were dereddened using the extinction curve of
\citet{car89}.

\section{Analysis}

\subsection{Hot Spots}

\begin{figure}
\plotone{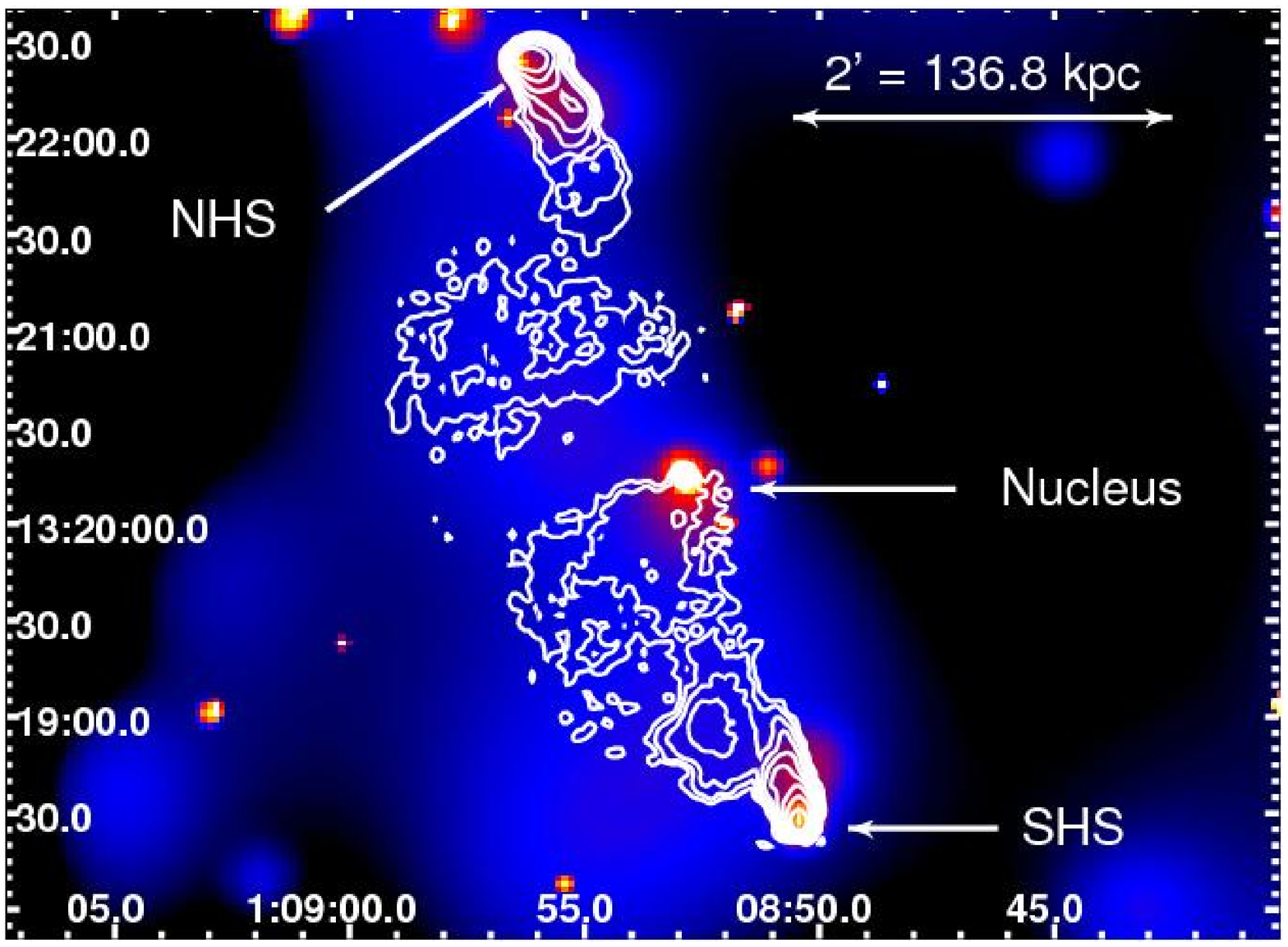}
\caption{Adaptively smoothed Chandra/ACIS-S image of 3C 33 in the 0.5-2.0 keV band
with 1.4 GHz radio contours ($4\farcs0$ resolution) overlaid.  The positions of the
active nucleus and the north (NHS) and south (SHS) hot spots are labeled.}\label{3c33cs}
\end{figure}

\begin{figure}
\plotone{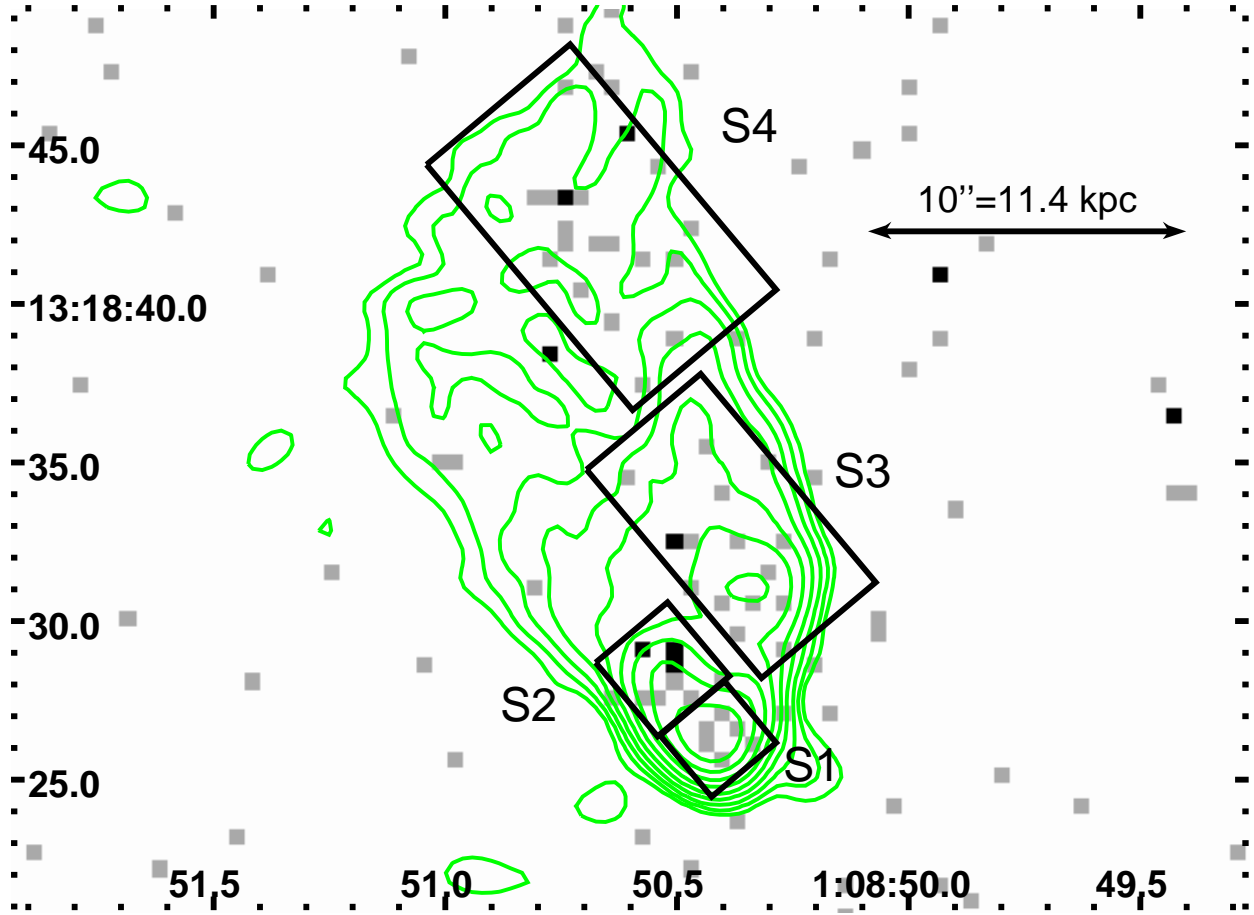}
\caption{Unsmoothed Chandra/ACIS-S image of the SHS in the 0.5-5.0 keV band
with 4.9 GHz radio contours ($1\farcs37 \times 1\farcs29$ resolution)
overlaid.  Regions S1 through S4 are also shown.
Contour levels correspond to 2.0, 3.7, 6.9, 12.9, 23.9,
44.4, 82.6, 153, 285, and 531 mJy beam$^{-1}$.}\label{shsraw}
\end{figure}

\begin{figure}
\plotone{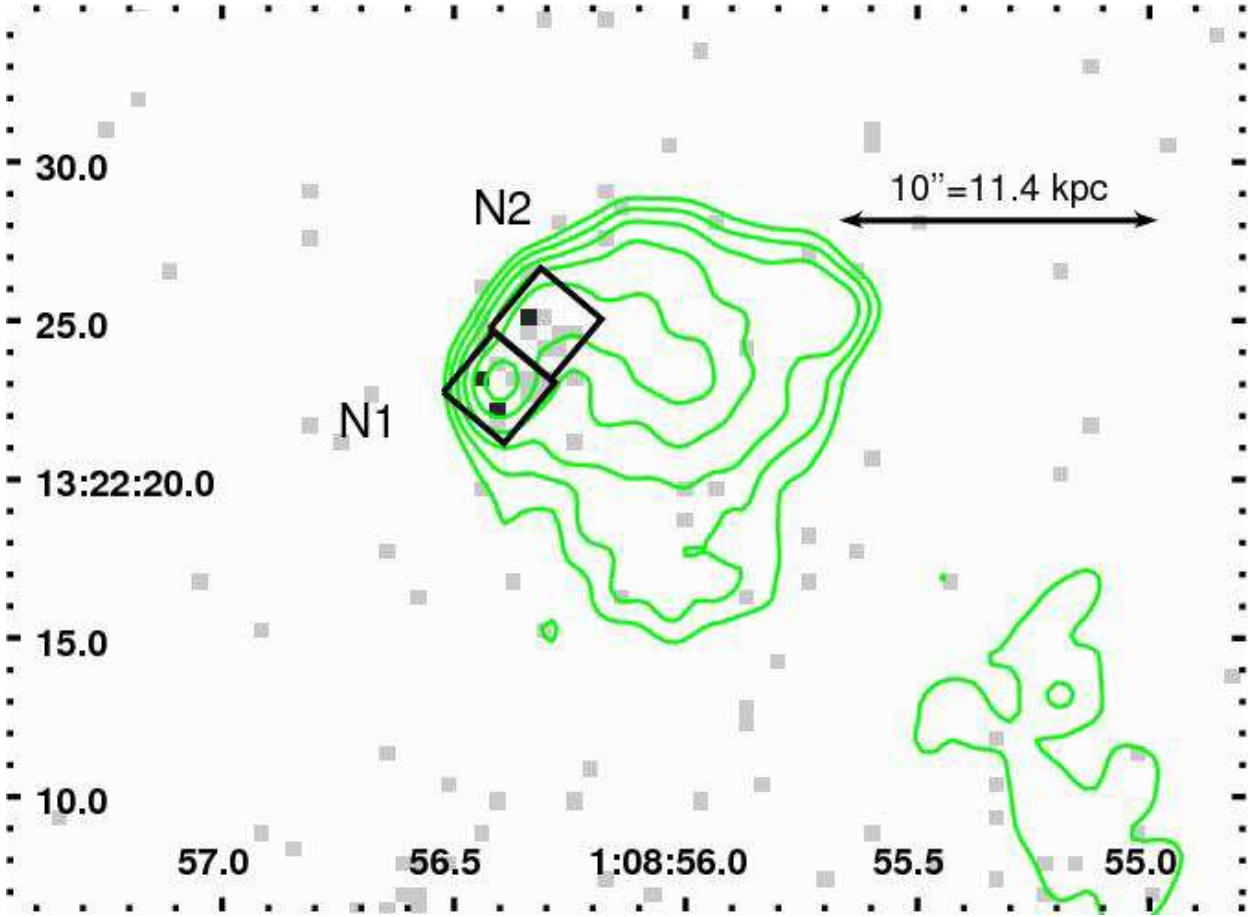}
\caption{Unsmoothed Chandra/ACIS-S image of the NHS in the 0.5-5.0 keV band
with 4.9 GHz radio contours ($1\farcs37 \times 1\farcs29$ resolution)
overlaid.  Regions N1 and N2 are shown, region N3 lies to the southwest.
Contour levels are the same as Figure~\ref{shsraw}.}\label{nhsraw}
\end{figure}

An adaptively-smoothed Chandra/ACIS-S image of 3C 33 with 1.4 GHz radio contours
overlaid is shown in Figure~\ref{3c33cs}.  The positions of the nucleus
and the two hot spots (NHS and SHS) are labeled.
Unsmoothed {\em Chandra}/ACIS-S images of the SHS and NHS in the 0.5-5.0 keV band
with 4.9 GHz radio contours overlaid are shown in Figures~\ref{shsraw} and
\ref{nhsraw}, respectively.
We have divided the SHS and the NHS into four and three regions, respectively,
for analysis.  The regions S1-S4 and N1-2 are shown in the two figures and
summarized in Table~\ref{regstab}.
Two additional regions, labeled DN1 and DS1,
coincident with diffuse X-ray emission between the nucleus and the radio hot spots
are also contained in Table~\ref{regstab}.
The 1 keV flux density is given for each region assuming a power law
spectrum with photon index 2 and Galactic absorption, consistent
with the spectral fits described below.

\begin{figure}
\plotone{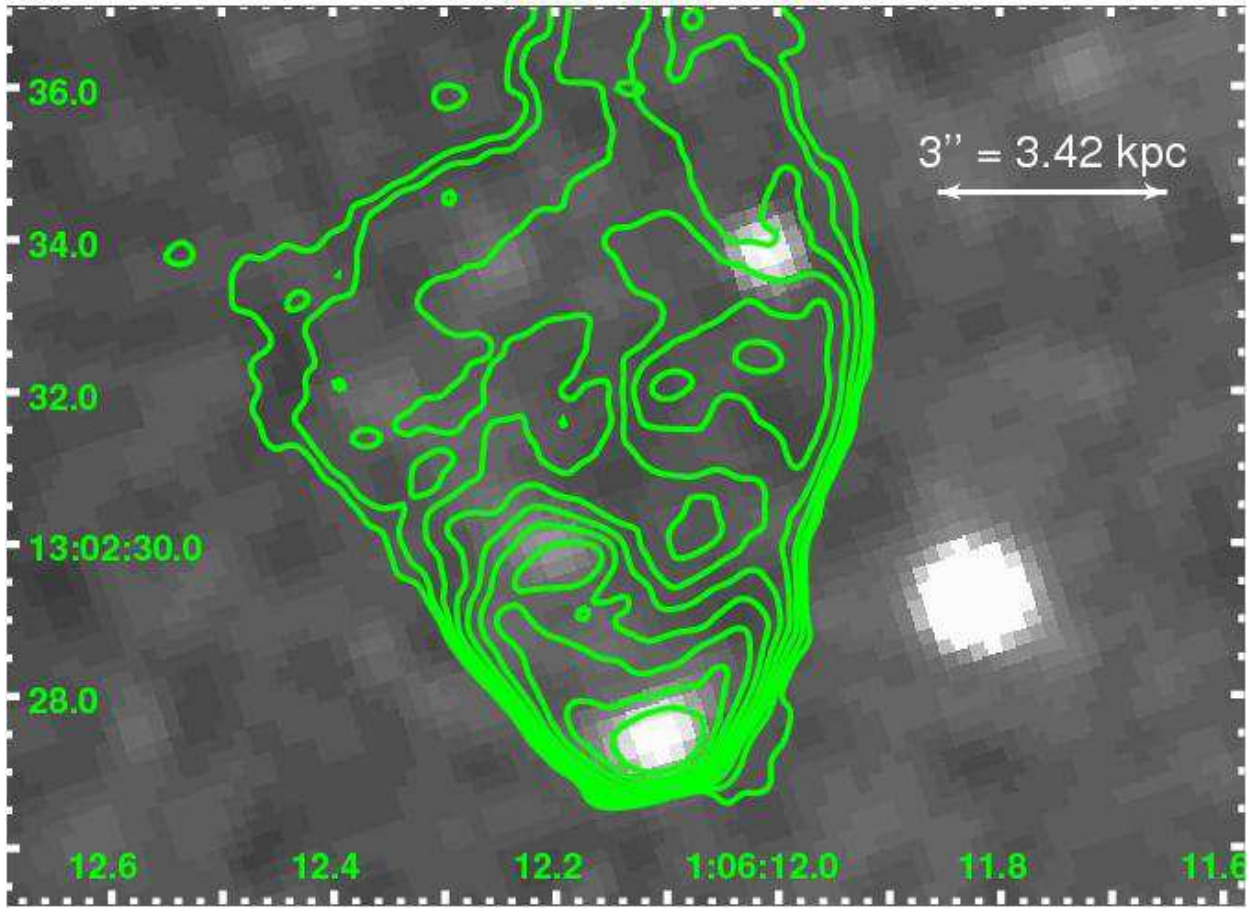}
\caption{WFPC2 (F702W filter) image of south hot spot with
4.9 GHz radio contours ($0\farcs43 \times 0\farcs38$ resolution) overlaid.}\label{shst}
\end{figure}

In both the NHS and SHS, spatially-resolved X-ray emission is detected
coincident with the compact radio peaks (regions S1 and N1).
In the SHS, X-ray emission is also detected
closer to the nucleus than the hot spot 
to the northeast (region S2), and along the entire northwest
half of the lobe (regions S3 and S4).
An {\em HST}/WFPC2 image of the SHS with 4.9 GHz radio contours overlaid
is shown in Figure~\ref{shst}.  An optical peak coincident with the
radio peak (S1) is observed, as is a weaker feature $\sim$2$''$ to the northeast (S2).
Both of these features have X-ray counterparts.
There is no apparent optical emission from S3 or S4.
In the NHS, a second region of X-ray emission
is detected in the NHS to the northwest of the radio peak (labeled region N2).
There is also a tail of diffuse X-ray emission behind (i.e. toward the
nucleus) N1 and N2.  This can be seen in Figure~\ref{3c33cs} and has been
labeled N3 in Table~\ref{regstab}.

Before proceeding, we must clarify an important point of terminology.  Historically
the bright radio features of 3C 33 (in their entirety) that we have termed the NHS and SHS 
in this paper have been generally been referred to as `hot spots' because they
are relatively compact.  This nomenclature is not entirely correct, however, as they contain
both compact features and resolved, diffuse emission behind the
compact features.  Neither are they `lobes', however,
such as are seen in the canonical FR IIs Cyg A and 3C 98.
For historical continuity, we will continue to use the acronyms NHS and SHS
or the term `hot spot' to refer to the entire radio bright features,
but we will use the definitions and nomenclature of \citet{lea97} when discussing
our individual regions.  By these criteria, S1 and N1 are primary hot spots.  They
are compact and are reasonably associated with the termination of the jet.
Regions S2 and N2 are secondary hot spots.  They are clearly hot spots
in a morphological sense, although they are not the termination points
of the jet.  Finally, regions S3, S4, and N3 are considered to be part of the `hot spot
complex', although they are not hotspots.
We will also use the term `hot spot' to refer to compact features of
FR IIs and radio quasars that are believed to be the termination of the jet.
The term `hot spot' will not be used in connection with individual regions
that are clearly not compact (e.g. S3, S4, and N3).

There are too few counts to fit the X-ray spectra of each region of the NHS
and the SHS individually (typically 10-15 source counts in
each region), so we combined all the counts in S1-S4 and N1-N3
to create two spectra.  We fit each spectrum with
an absorbed power law model, with the absorption fixed at the Galactic
value to reduce the number of free parameters.
The best fit photon indices are 1.8$\pm$0.6 and 2.2$\pm$0.8 for the SHS and NHS,
respectively.  The quality of these data are not sufficient to
discriminate between emission models (i.e. synchrotron, thermal, or
inverse Compton scattering) on the basis of spectra, although
we reject the thermal model for reasons detailed below.
The unabsorbed X-ray luminosities of the SHS and NHS
are $\sim$1.6 and $\sim$0.5$\times$10$^{41}$ ergs s$^{-1}$, respectively,
in the 0.5 to 5.0 keV bandpass.

\begin{figure}
\plottwo{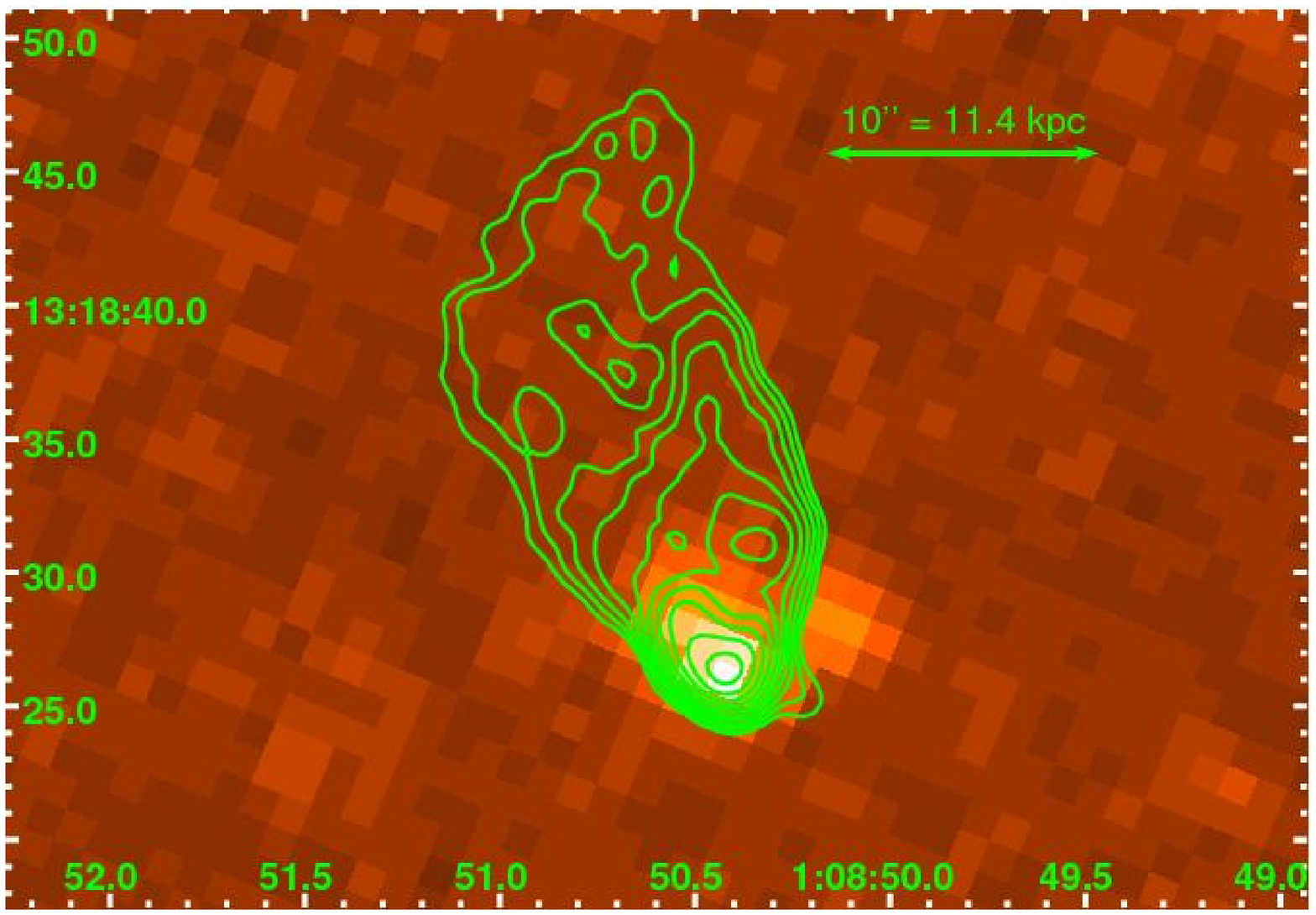}{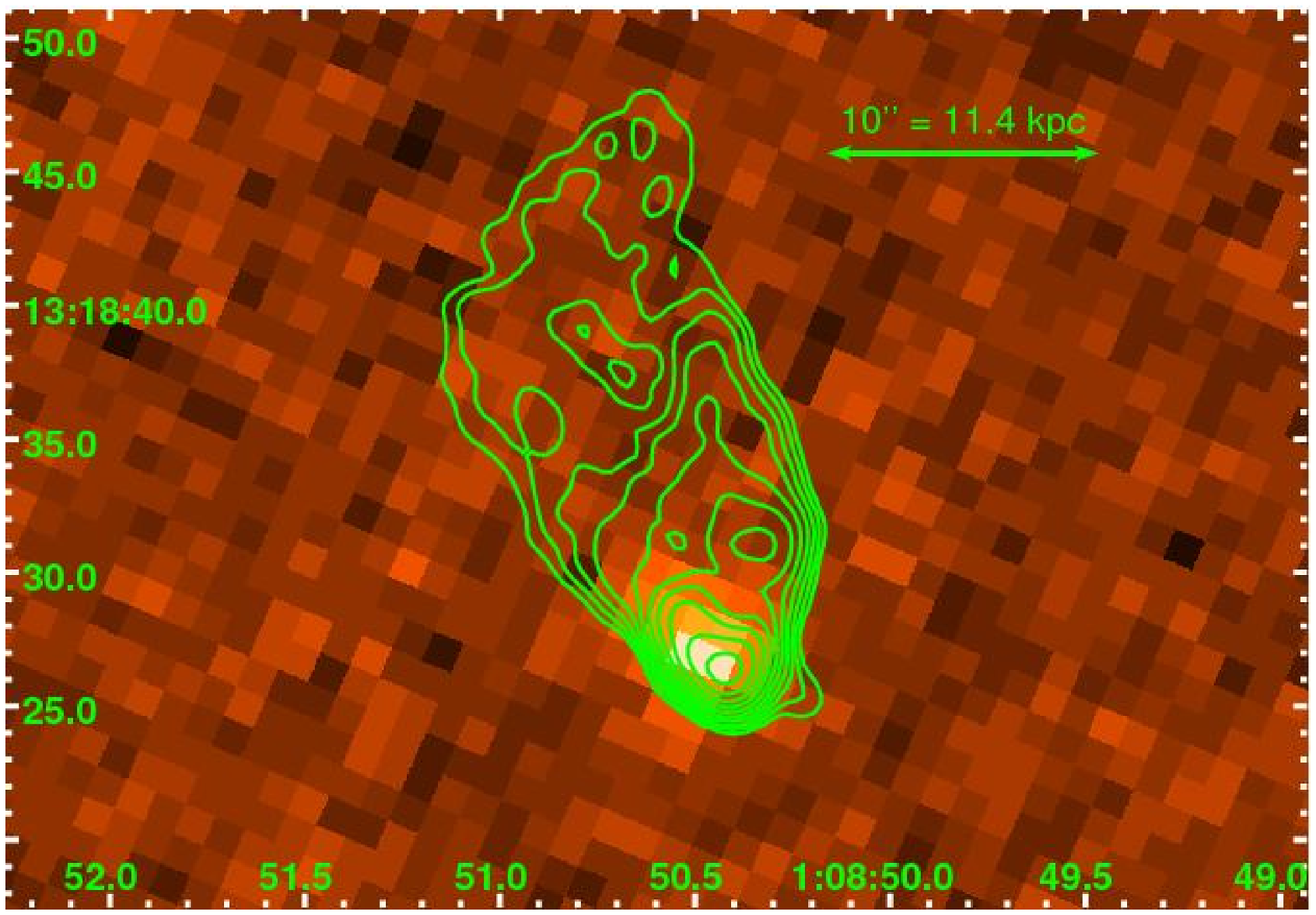}
\caption{{\em Spitzer}/IRAC images of the SHS at 4.5 (left) and 8.0 (right)
$\mu$m with 4.9 GHz radio contours ($1\farcs37 \times 1\farcs29$
resolution) overlaid.  The radio contours are logarithmically
spaced between 3.0 and 400.0 mJy beam$^{-1}$.}\label{spitzer}
\end{figure}

\begin{figure}
\plotone{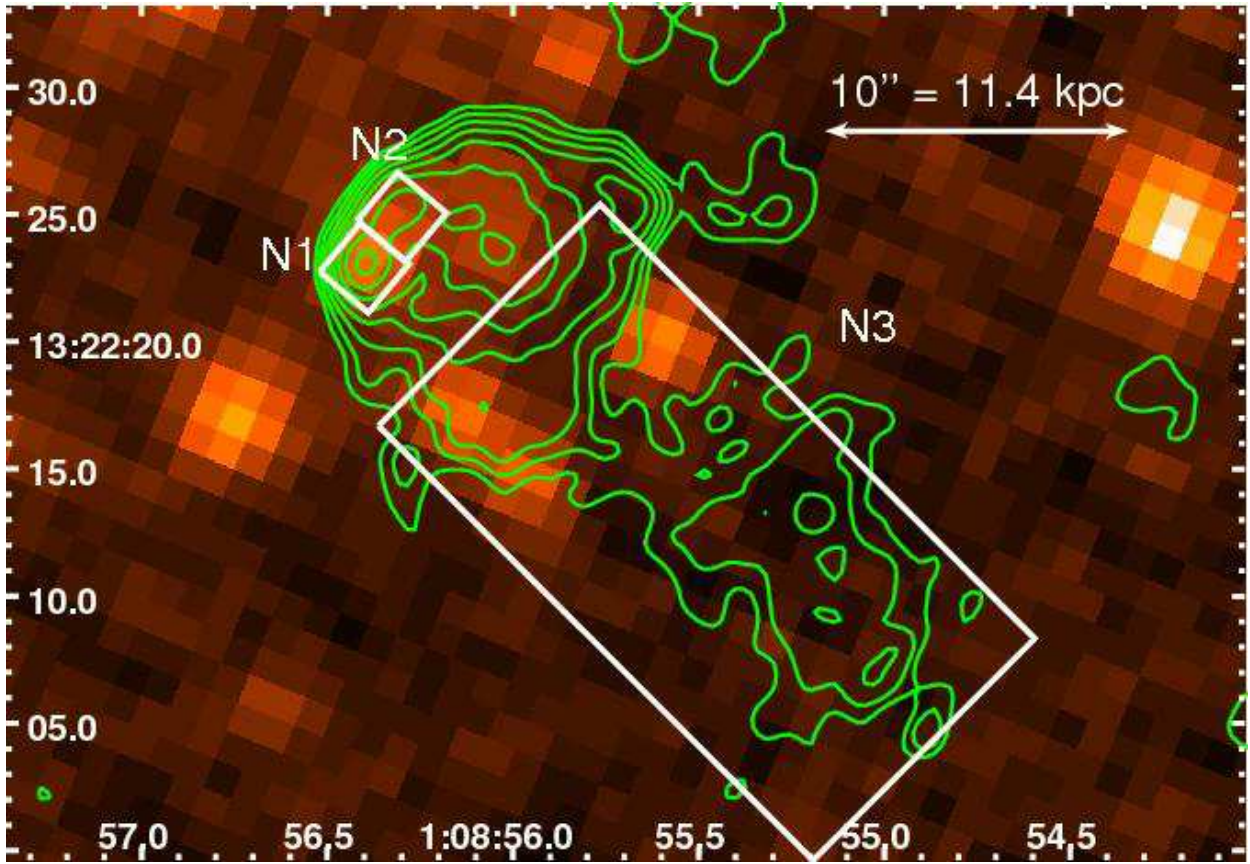}
\caption{Smoothed (Gaussian - $\sigma$=1.2$''$) {\em Spitzer}/IRAC image of the NHS 
at 4.5 $\mu$m with 1.4 GHz radio contours ($4\farcs0 \times 4\farcs0$
resolution) overlaid.  The positions of N1 and N2 are
also shown.}\label{nhsirc}
\end{figure}

The morphology of the SHS in the infra-red in two {\em Spitzer}/IRAC
bands, 4.5 and 8.0 $\mu$m, is shown in Figures~\ref{spitzer}.
At the shortest IRAC wavelengths (Figure~\ref{spitzer} left),
region S1 dominates the IR emission of the SHS,
although S2 is clearly detected.  There is little or no emission from
S3 or S4.  At longer wavelengths (Figure~\ref{spitzer} right),
the fluxes from S1 and S2 are roughly equal.  Thus the flux density
spectral index (and therefore the electron energy distribution index) is
considerably steeper for S2 than S1.
The morphology of the NHS in the IR at 4.5 $\mu$m, shown in
Figure~\ref{nhsirc}, is very different than that of the SHS.
There is a bright IR peak coincident with region N1, and a $\sim$8$''$
long tail of emission to the west that follows the radio contours through
region N2.  The limited spatial resolution of {\em Spitzer} ($\sim$1.2$''$) 
prevents us from determining whether this emission is continuous or composed of
small (kpc-scale) knots.  If the radio peak (N1) represents 
the termination point of the jet, the detection of this filament of
IR emission perpendicular to the hot spot/jet axis suggests that
there is particle acceleration (and therefore large velocities) perpendicular
to this axis as well.
This is qualitatively consistent with the `primary + splash model' for the flow
pattern of the NHS proposed by \citet{rud90}.

The angular resolution of {\em Spitzer} is such that the S1 and S2
components of the SHS, and the N1 and N2 components of the NHS, are
separated in IRAC channels 1 and~2 (3.6 and 4.5 $\mu \rm m$), but
blended to a greater or lesser extent at the other wavelengths. The
fields are further complicated by the presence of nearby foreground stars. 
To extract the flux densities of the separate components we therefore
fitted point sources at the approximate locations of the components
(leaving some freedom in the coordinates to allow for a misalignment
of the radio and IR coordinate systems), and at one or two
contaminating stars in the hot-spot fields. Cross-checks
of the derived flux densities were made using aperture photometry,
where possible. Errors on the flux densities of the components were
derived from the fits taking account of the residuals in the areas
fitted and the number of parameters being fitted, and include the
systematic errors on the flux density scale expected in the
different bands.

At 24$\mu$m this degree of fitting was not possible, because
the hot spot components are severely blended by the
diffraction-limited point spread of {\em Spitzer}.  At this band we therefore
fitted the best flux densities using defined locations for the hot
spot components, and compared the results with the fraction of the
total flux densities for the hot spot regions that would be expected
given their 3.6 to 8 $\rm \mu m$ spectra. Generally the results agree,
though the errors from the spectral extrapolation constrained by the
total flux densities of the hot spots are smaller, and so the latter
results were adopted. 

The results for the IR flux densities, as well as
radio flux densities, of all seven regions
(S1 through S4 and N1 through N3) are shown in Table~\ref{fdsumtab}.
For completeness, the optical, UV, and X-ray flux densities (or
upper limits) for S1 are shown in Table~\ref{fdsumtaboptx}.
The optical points are taken from either archival {\em HST} images
or previously published values \citep{mei86,cra92}.
Regions S3, S4, and N1 through N3 were not detected in any
optical or UV band.

We have constructed radio through X-ray flux density distributions for six regions.
These are shown in Figures~\ref{s12sed},~\ref{s34sed},~\ref{n12sed}
for the S1/2, S3/4, and N1/2 pairs, respectively.
All error bars are 1$\sigma$ statistical uncertainties.  If the error bars
are not visible, they are smaller than the data points.
The upper limits on the IRAC fluxes shown in Figure~\ref{s34sed}
are 3$\sigma$ limits.  The solid lines are representative single injection
synchrotron models, and the dotted and dashed lines are the estimates 
of the minimum energy IC/CMB and SSC models,
respectively.  The details of these spectral models, as well as implications regarding
X-ray emission mechanisms, are presented in Section~4.

\subsection{Nucleus}

\begin{figure}
\plottwo{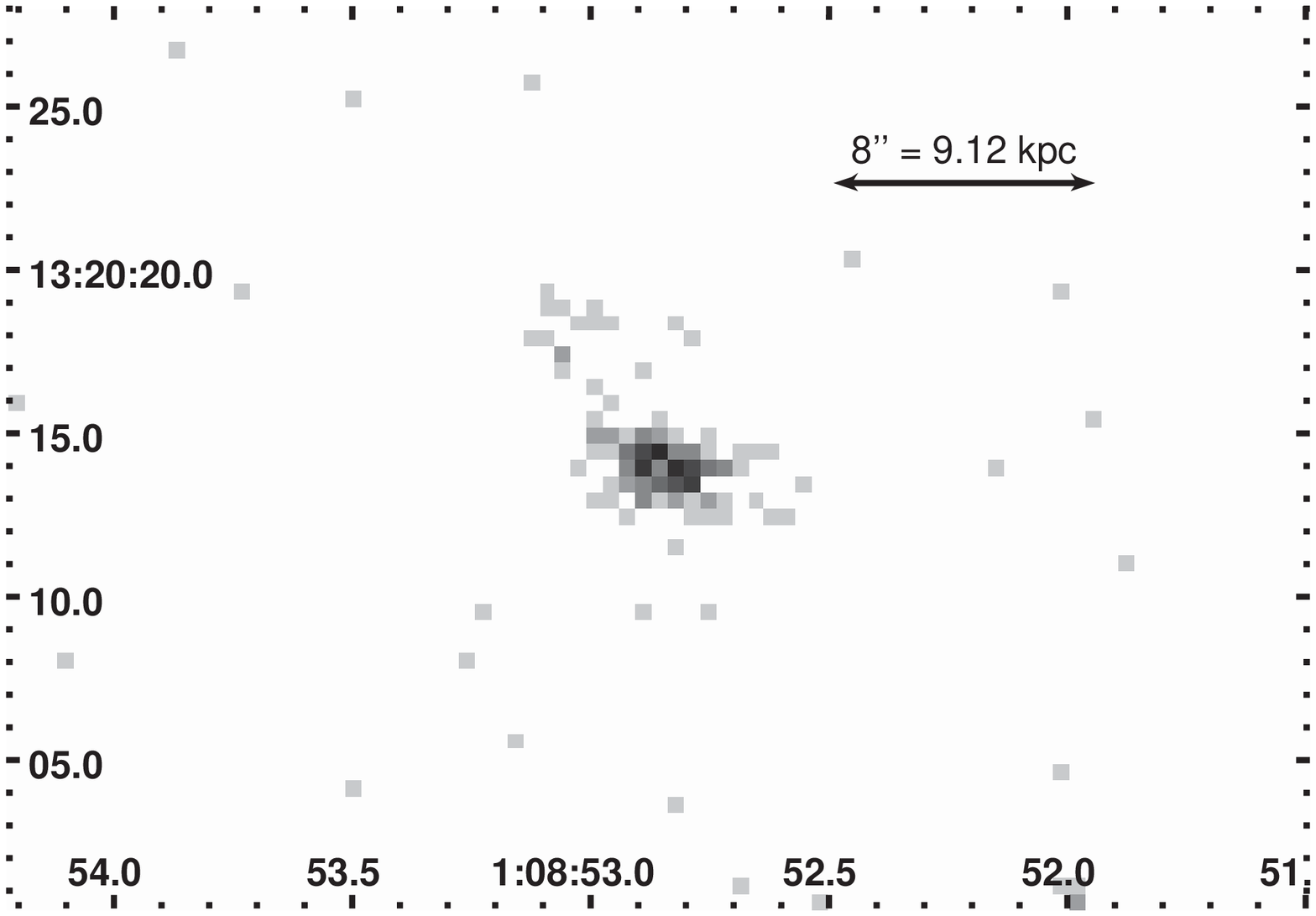}{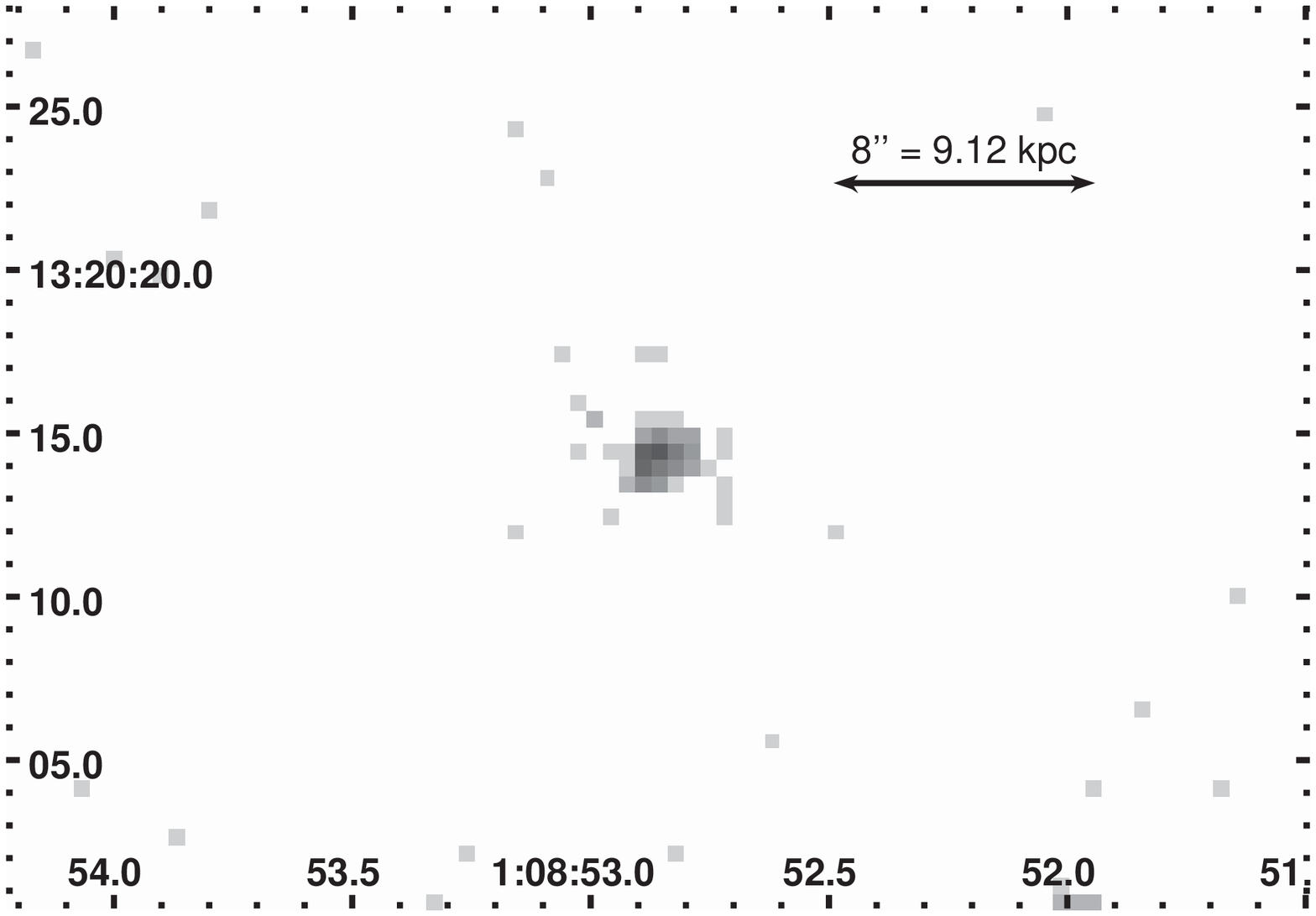}
\caption{Unsmoothed {\em Chandra}/ACIS-S X-ray image of the 3C 33 nuclear region in the
0.3-0.7 keV (left) and 1.5-2.0 keV (right) bands at 
full resolution (1 pixel = 0.492$''$).}\label{nimages}
\end{figure}

X-ray images of the nuclear region of 3C 33 in the 0.3-0.7 (soft) and 1.5-2.0
(hard) keV bands are shown in Figure~\ref{nimages}.  In the hard
band, the nucleus is unresolved and consistent with a point source.  In the soft
band the structure is clearly extended, with the emission
elongated along the northeast/southwest axis.  This extension is not
the frame transfer streak (FTS) as it
lies roughly along the ENE/WSW axis through the nucleus.
The position angle of the observed extension is rotated relative to this
axis by approximately 30$^\circ$.  In addition, the nucleus contains only
$\sim$2000 counts, so $\sim$30 counts are expected over the entire $\sim$8$'$
length of the FTS.  {\em HST} observations of the
host galaxy show a similar elongation in the stellar distribution suggesting
that the extended X-ray emission is from thermal gas.  
There is an additional spatially extended X-ray component $\sim$5$''$ (5.7 kpc) to the northeast.
It is unclear whether this is thermal emission from an extended gaseous
component or X-ray emission from a nuclear jet.  There is no evidence of
a kpc scale jet in any of our radio maps.

The spectrum of the nucleus was extracted from a 
source-centered circle of radius 10 pixels (4.92$''$), with background sampled 
from a surrounding annulus of inner 
radius 10 pixels and outer radius 20 pixels (9.84$''$).
The {\em Chandra} count rate from the nucleus is $\sim$0.05 cts s$^{-1}$, 
indicating little pileup.  
Initially we fitted the spectrum of the nucleus with a three-component model: an
absorbed power law, a Gaussian line representing emission from neutral
Fe K$_\alpha$, and a thermal (the XSPEC Astrophysical Plasma Emission Code model, 
or simply APEC) component.  
The fit was poor ($\chi^2_\nu$=19.9 for 26 degrees of
freedom) with large residuals between 1 and 3 keV, so we investigated two alternative
models.  

\begin{figure}
\plottwo{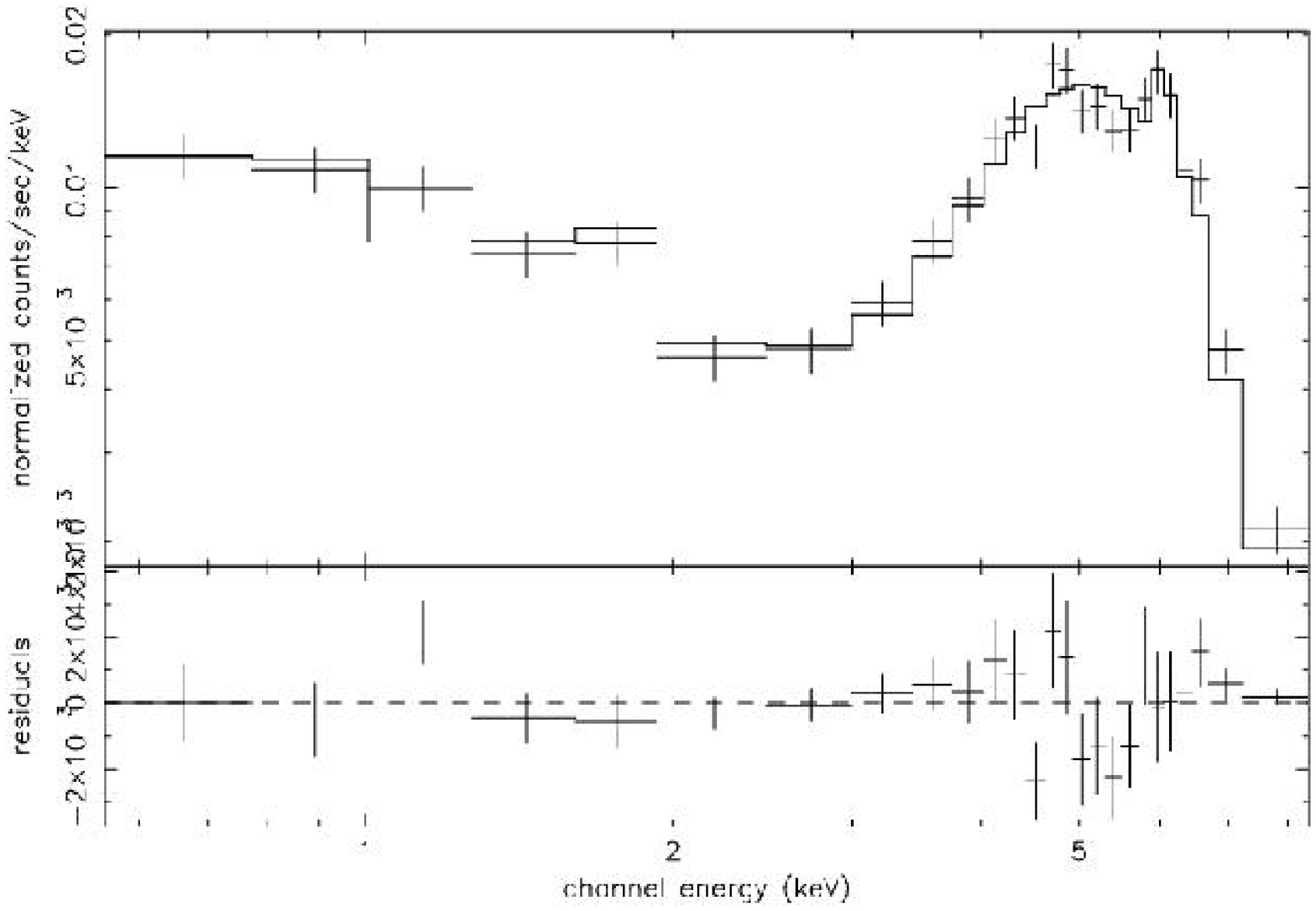}{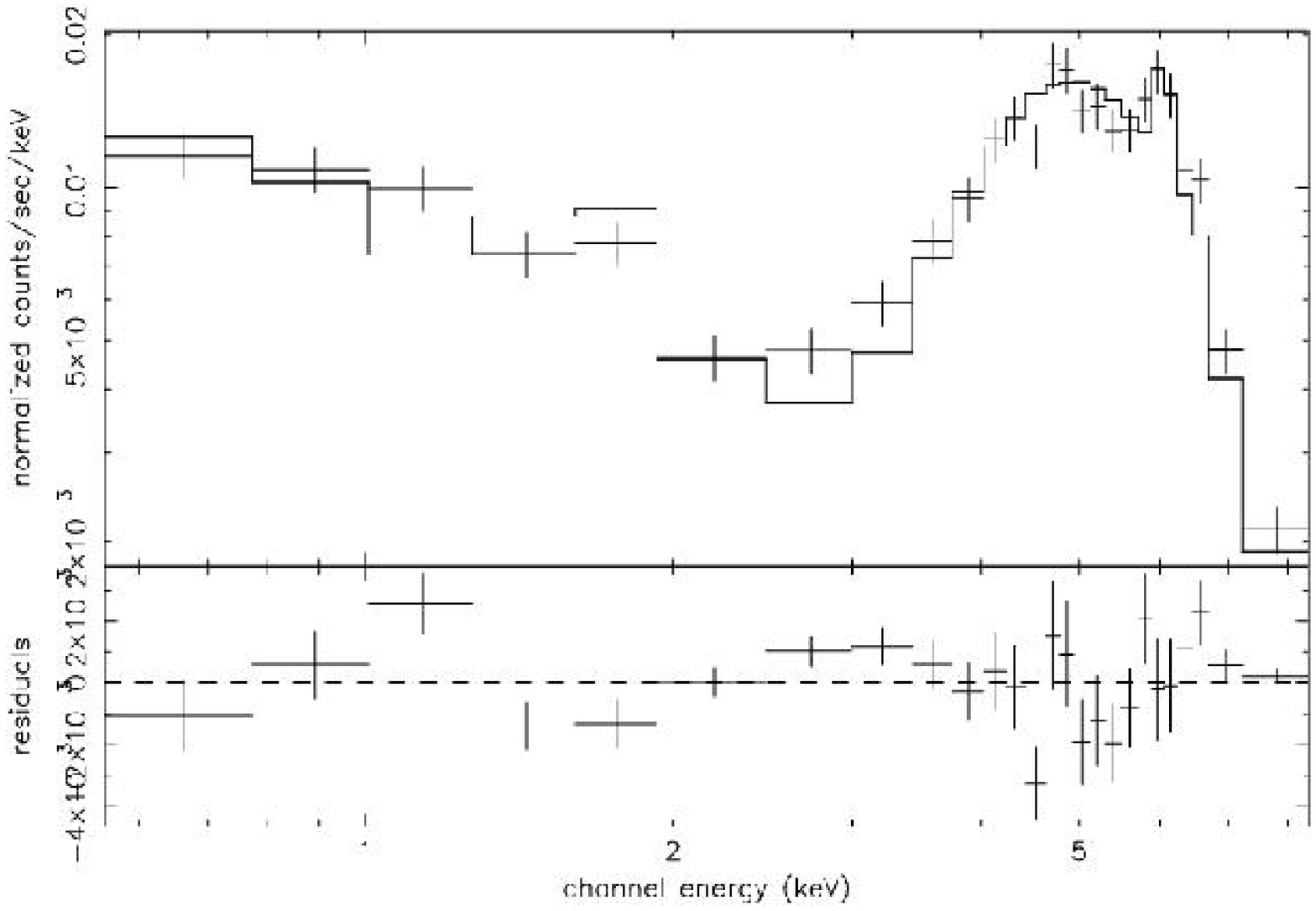}
\caption{Best-fit X-ray spectrum and residuals of the nucleus of 3C 33.  The model
(Model 1) shown in the left figure consists of a single heavily absorbed power law, a reflection
component, a Gaussian line at $\sim$6.4 keV, and a thermal
(APEC) component.  The model shown in the right figure (Model 2) consists of a single 
heavily absorbed power law, a second,
less absorbed, power-law component, a Gaussian line at $\sim$6.4 keV, and a thermal
(APEC) component.  The best-fit values of the parameters and
uncertainties are contained in Table~\ref{spectab}.}\label{specfit}
\end{figure}

Model 1 consisted of a four component model, the three components listed in the previous paragraph
plus a reflection component.
The reflection model has been used to fit the X-ray spectra of Seyfert galaxies,
but has rarely been found to provide an adequate description of the
X-ray spectra of radio-galaxy nuclei.
The X-ray spectrum with the best-fit four-component reflection
and two power-law component models and residuals are shown in 
the left and right plots of Figure~\ref{specfit}, respectively.
Model 2 also consisted of the three components 
listed in the previous paragraph plus a second,
less absorbed power law representing emission from a parsec scale jet.  Such a
model has been shown to be a good description of emission from the nuclei of
FR II radio galaxies \citep{evans06}.  
A summary of best-fit parameters and uncertainties for both
models is contained in Table~\ref{spectab}.
The APEC component in the model that includes
a reflection component required abundances of Mg and Si that were larger than the
Solar value and a sub-Solar value for Fe.
This is probably not physical and may be indicative of systematics
in the model such as multi-temperature gas in the ISM or more
complex reflection geometry than a uniformly illuminated slab.
We note that the large column density toward the primary power law in both
spectral fits is consistent
with the expectations from unified models for a narrow-line radio
galaxy, as is generally seen in X-ray observations of large samples of such 
sources (e.g. \citealt{sam99,bel06,evans06,mjh06}).  This strongly supports the idea
that the jet axis is close to the plane of the sky and that relativistic beaming
is not important in modifying the appearance of the hot spots.
Finally, we mention that we fitted the {\em XMM-Newton} spectra (both MOS cameras and PN)
with the same models and find that the fits are consistent with the
{\em Chandra} results.

\subsection{Large Scale Diffuse Emission and Environment}

\begin{figure}
\plotone{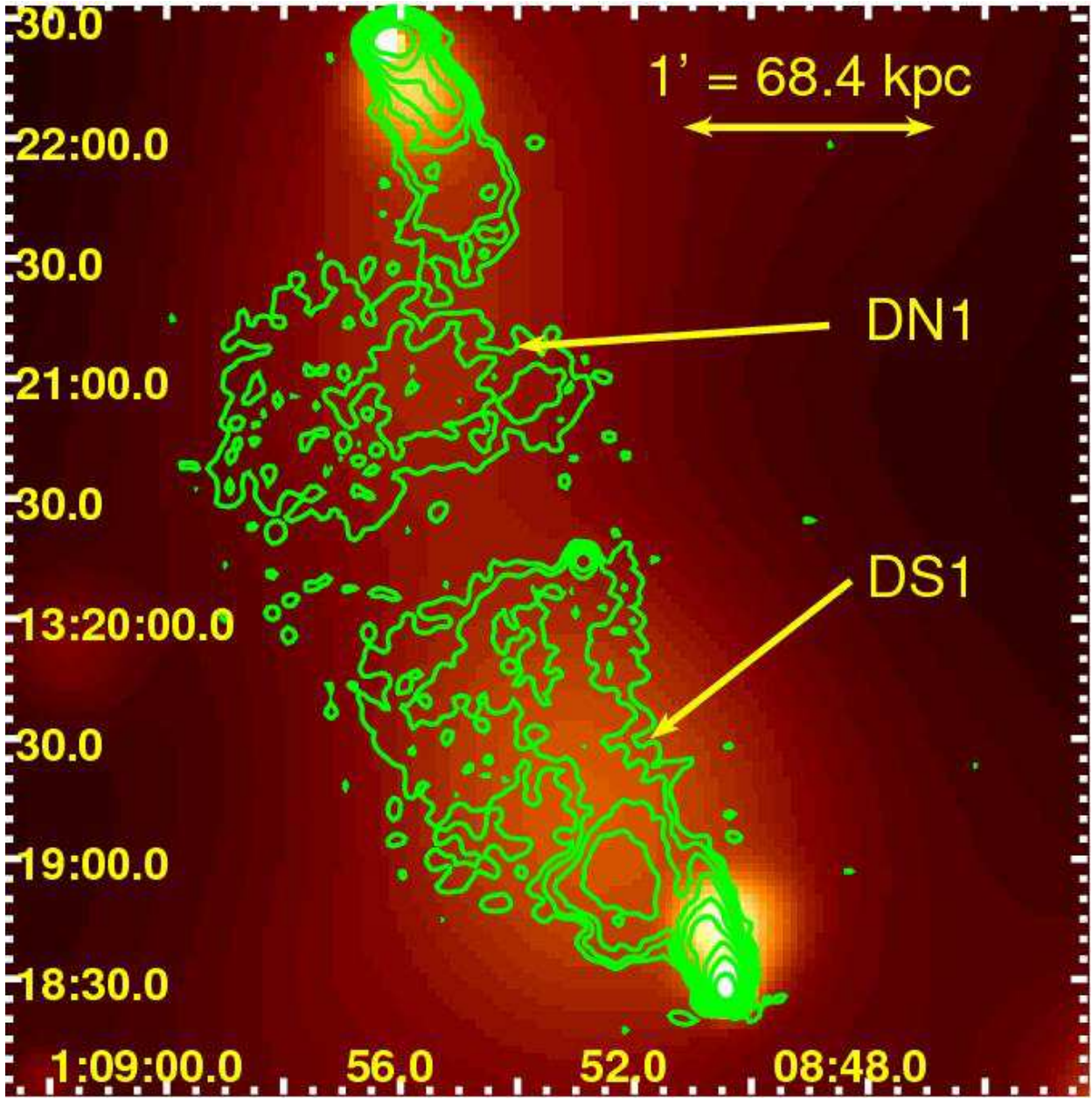}
\caption{Adaptively smoothed {\em Chandra}/ACIS-S X-ray image in the
0.5-2.0 keV band with all point sources other than the X-ray hot
spots removed.  1.5 GHz radio contours are overlaid.  The two regions
of diffuse X-ray emission, DN1 and DS1, associated with the low surface
brightness radio emission between the hot spots and nucleus have been
labeled.  We attribute this X-ray emission to inverse-Compton scattering
of CMB photons off the relativistic electrons.}\label{tail}
\end{figure}

Diffuse, low surface brightness emission unrelated to either the hot
spots, the central galaxy, or the active nucleus is visible in
Figure~\ref{3c33cs}. This emission is more clearly shown in
Figure~\ref{tail}, an adaptively smoothed X-ray image in the 0.5-2.0
keV band with all point sources other than the NHS and SHS removed.
Two regions of diffuse X-ray emission have been labeled DN1 and DS1.
This emission is generally aligned along the NE/SW axis of the radio
structure, and extends more than 100 kpc. There is a rough spatial
correspondence between this X-ray emission and the diffuse radio lobe
emission between the hot spots and the nucleus. The X-ray fluxes of
regions DN1 and DS1 are 4.5 and 6.2$\times 10^{-15}$ ergs cm$^{-2}$
s$^{-1}$ assuming a power-law spectrum with photon index 1.5 and
Galactic absorption: this corresponds to 1-keV flux densities of 1.3
and 1.8 nJy. The combined spectrum of the two regions is
well fitted with a power-law model of photon index 1.5$\pm$0.7
(90\% confidence) and Galactic absorption. 
We also fitted the spectrum with an absorbed APEC model with the absorption
fixed at the Galactic value and the elemental abundance fixed to
0.5 times Solar to reduce the number of free parameters.
The best fit temperature is 1.5$^{+1.9}_{-0.5}$ keV (90\% confidence for
one parameter of interest).  
We believe that this emission is non-thermal, but a purely thermal model
is not conclusively ruled out by the data (see Section 4.2 for complete
discussion).

\begin{figure}
\plotone{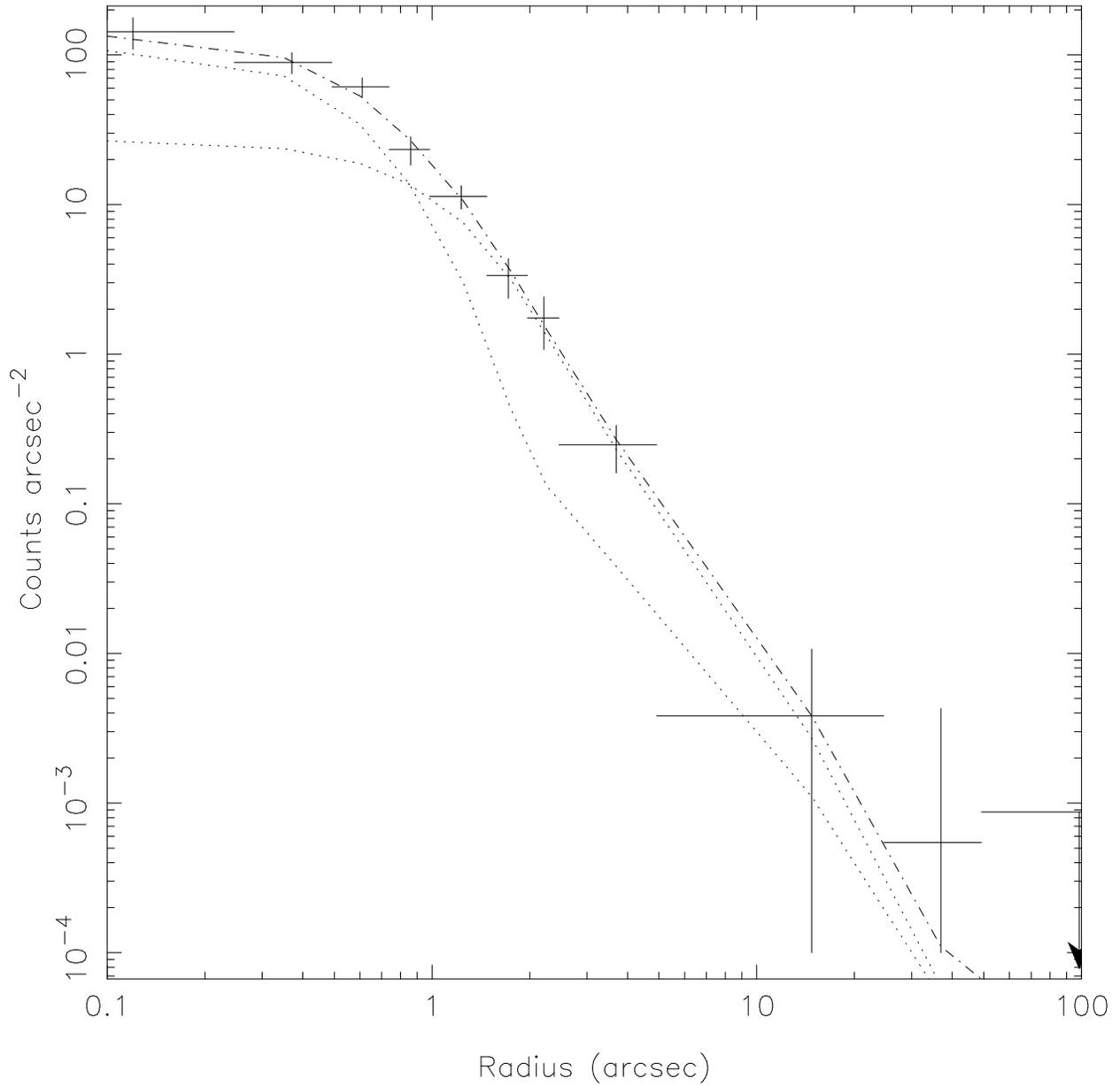}
\caption{Chandra radial surface brightness profile in the 0.5-2.0 keV
band.  The two dotted curves are the point source and beta-model profiles
convolved with the Chandra PSF.  The dot-dashed line is the sum of the
two dotted curves.}\label{sbprof}
\end{figure}

To investigate the properties of the undisturbed environment of 3C 33 on smaller 
(i.e. galactic) scales, we extracted a radial surface brightness profile from the {\it
Chandra} data in the 0.5 - 2 keV energy range, excluding the radio
lobe and hotspot regions.  This profile is shown in Figure~\ref{sbprof}.
The total number of counts (after background subtraction)
in the surface brightness profile is $140^{+51}_{-28}$ cts.
We modeled the {\it Chandra} point-spread
function using ChaRT and Marx. A point-source model was not a good fit
to the profile, with a clear detection of extended emission between 10$''$
and $\sim$100$''$. We therefore fitted a point-source plus $\beta$
model to the profile, although the parameters of the $\beta$ model
were poorly constrained.  Acceptable fits were obtained for $\beta$ 
values in the range of 0.67 to 0.9, with $r_c$$\sim$2$''$ ($\sim$2.3 kpc). 
We extracted a spectrum
in an annulus between 10 and 100 arcsec, excluding the radio lobes and
hotspots, and found that the 0.5 - 2 keV spectrum is consistent with a
thermal model of $k_BT$=1.8$^{+0.2}_{-0.6}$ keV (the spectrum above 2
keV is dominated by scattered AGN emission). 
This temperature is unphysically large for a relatively isolated
elliptical galaxy and perhaps suggests that even the gas in
the central 10 kpc of the host galaxy has been disturbed by the
passage of the powerful radio jets.  We used the {\em
XMM-Newton} data to investigate whether any emission on larger scales
was present; the data are consistent with the presence of the {\it
Chandra}-detected component, but the poor data quality meant that no
strong upper limits on larger-scale emission could be obtained.

We integrated the best-fitting $\beta$ model ($\beta = 0.9$, $r_{c}$
= 1.3$''$) to determine the total net counts from the environment,
which corresponds to a count rate of $\sim$7$\times$10$^{-3}$
cts s$^{-1}$ in the 0.5-2.0 keV band. Using the best-fitting temperature, this
corresponds to a luminosity of $1.5 \times 10^{41}$ ergs
s$^{-1}$ in the 0.5-2.0 keV band, or a bolometric luminosity of 
$\sim$3$\times$10$^{41}$ ergs s$^{-1}$. The luminosity-temperature 
relation for radio-quiet groups from \citet{cro05a} predicts a luminosity of
$\sim$4$\times$10$^{43}$ ergs s$^{-1}$ for groups with a temperature of
$\sim$1.8 keV, so that the 3C\,33 group appears to be roughly two
orders of magnitude less luminous than would be expected for its
temperature, unless both {\it Chandra} and {\em XMM-Newton} are
failing to detect a very large, smooth extended environment. A
temperature of $\sim$0.5 keV is predicted for our measured
luminosity. This suggests either that 3C\,33 (or previous generations
of outbursts from its AGN) has had a dramatic effect on the
surrounding group, or that powerful radio galaxies can form in 
extremely poor environments.

\section{Interpretation}

\subsection{X-ray Emission from Hot Spots}

\begin{figure}
\plottwo{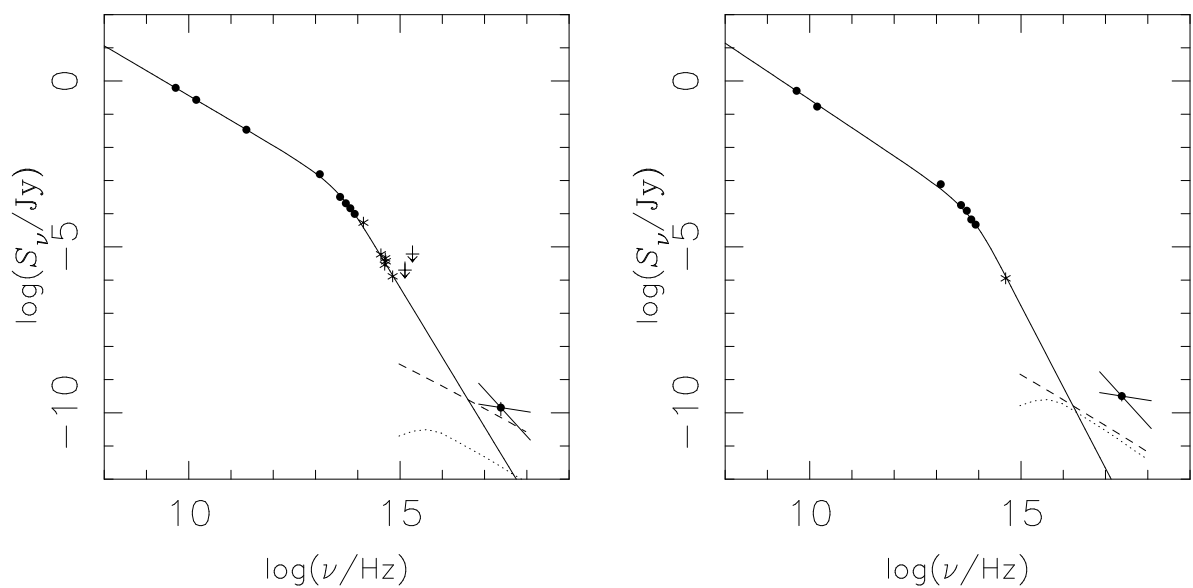}{./f7b.ps}
\caption{Radio through X-ray flux density distributions of regions
S1 (left) and S2 (right).  The continuous curve denotes the synchrotron spectrum,
and the dashed and dotted lines the estimates of the SSC and IC/CMB
X-ray fluxes, respectively. The optical and IR data for S1 include flux
densities taken from Meisenheimer et al. (1989) (as used by \citealt{mjh98}) as well as our new
radio, {\em Spitzer} and {\em HST} values.}\label{s12sed}
\end{figure}

\begin{figure}
\plottwo{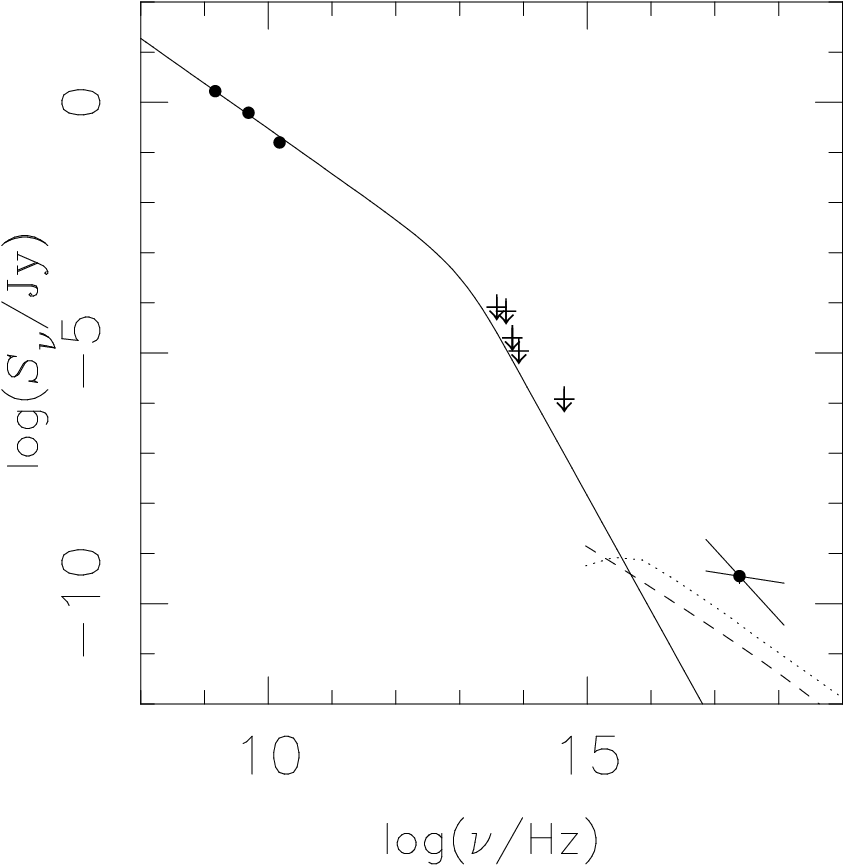}{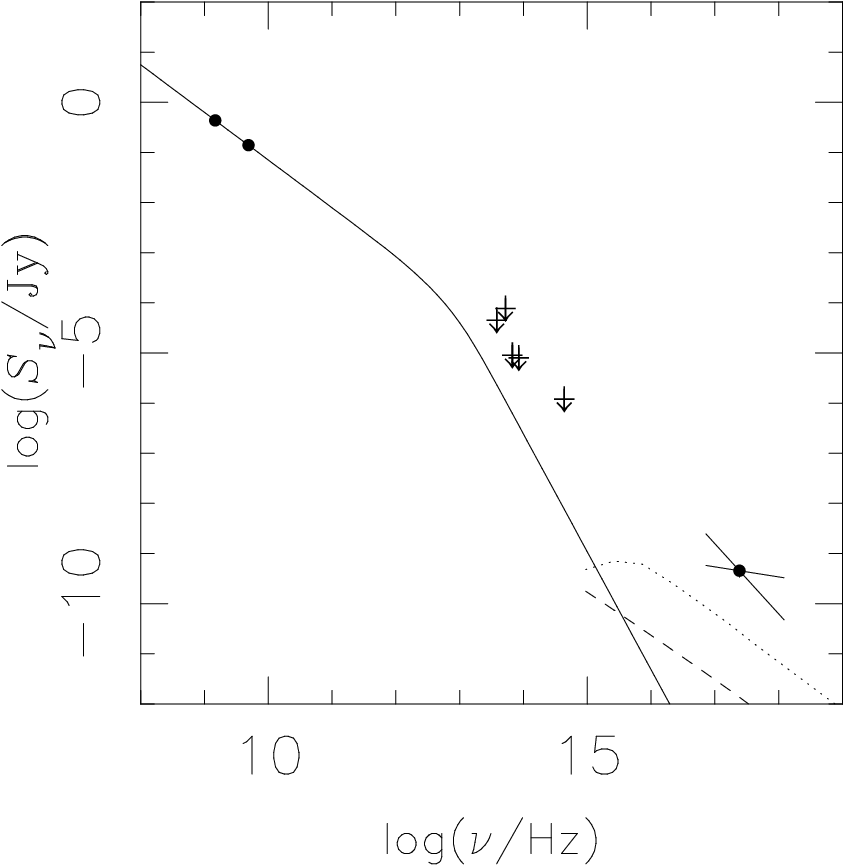}
\caption{Radio through X-ray spectral energy distribution of regions
S3 (left) and S4 (right).  The continuous curve denotes the synchrotron spectrum,
and the dashed and dotted lines the estimates of the SSC and IC/CMB
X-ray fluxes, respectively.}\label{s34sed}
\end{figure}

\begin{figure}
\plottwo{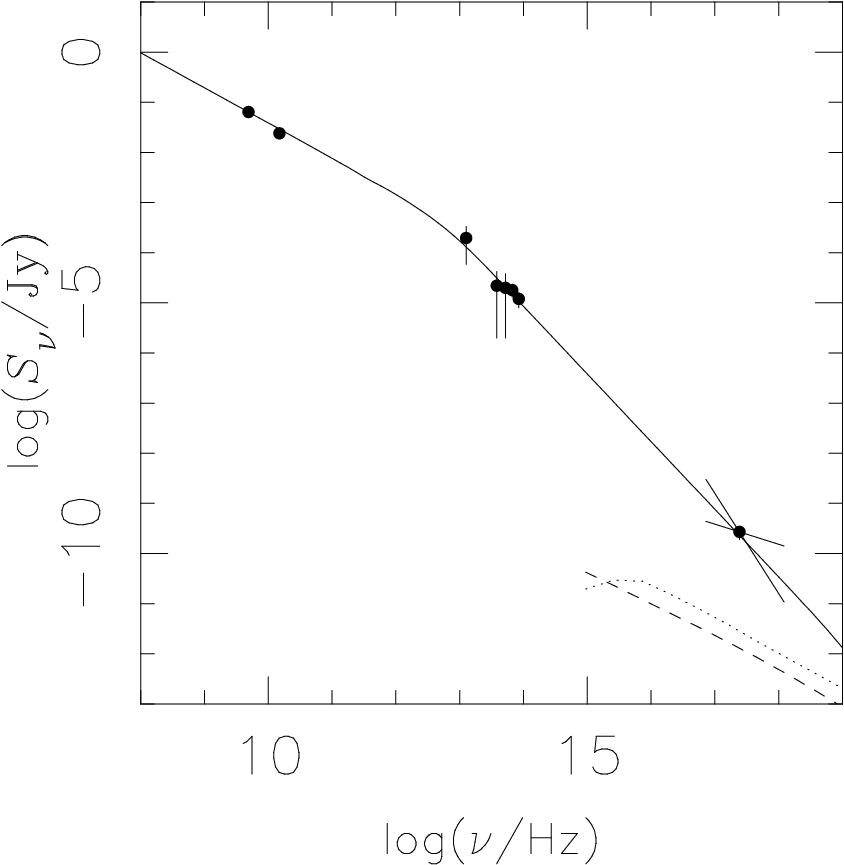}{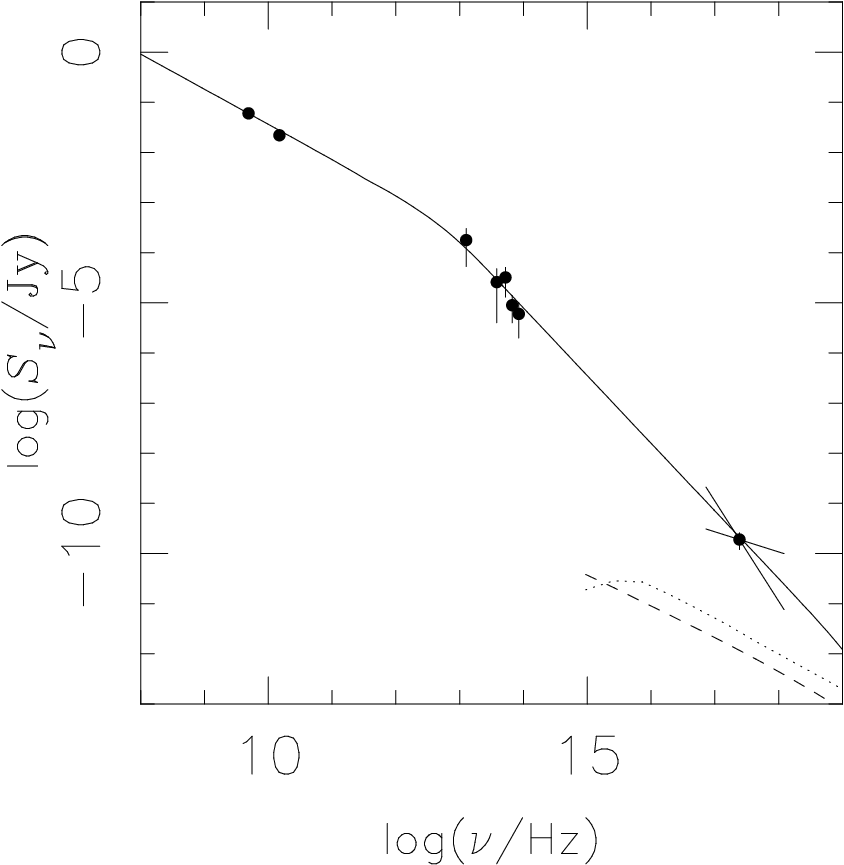}
\caption{Radio through X-ray spectral energy distribution of regions
N1 (left) and N2 (right).  The continuous curve denotes the synchrotron spectrum,
and the dashed and dotted lines the estimates of the SSC and IC/CMB
X-ray flux, respectively.}\label{n12sed}
\end{figure}

\begin{figure}
\plotone{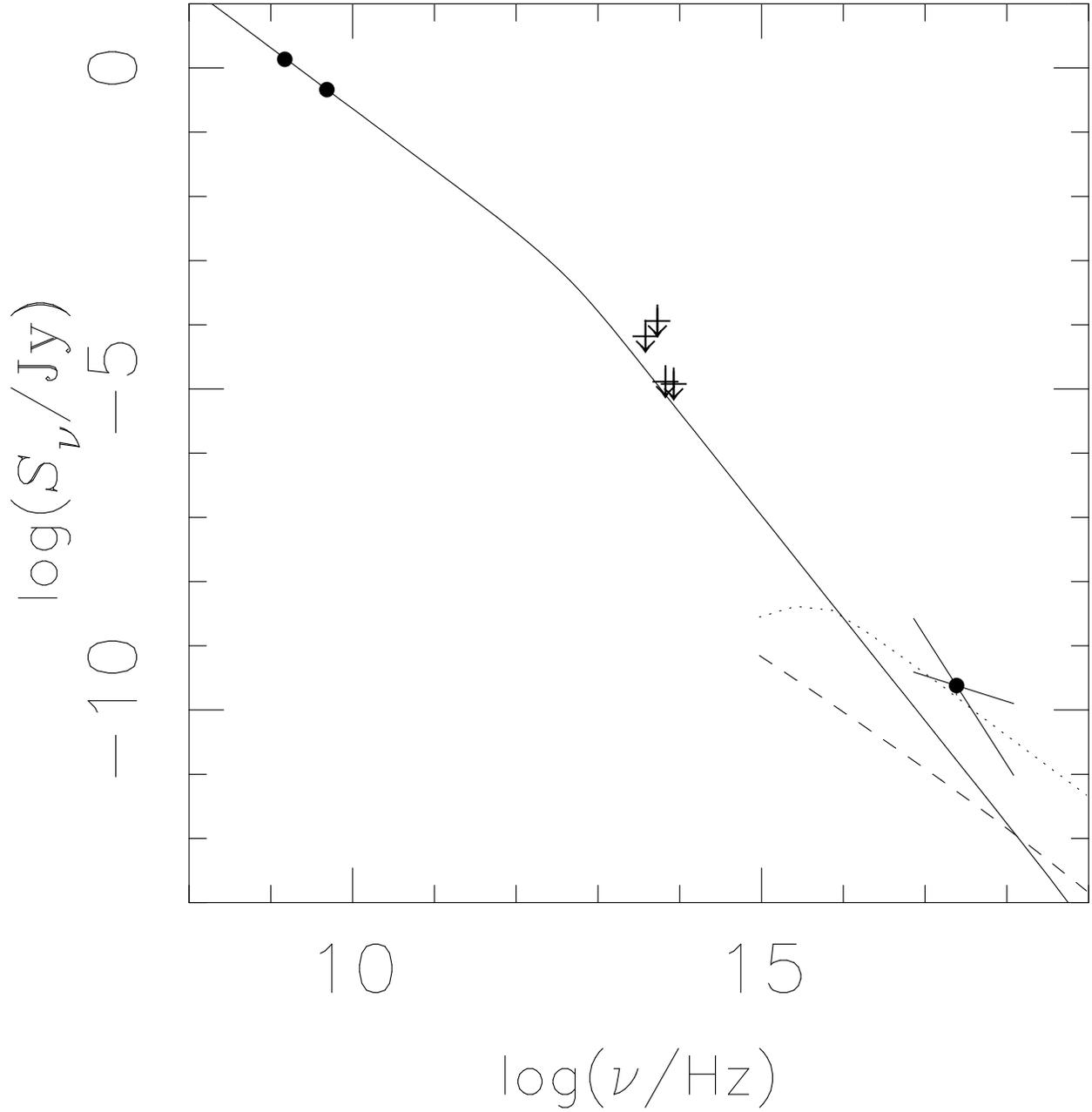}
\caption{Radio through X-ray spectral energy distribution of region
N3.  The continuous curve denotes the synchrotron spectrum,
and the dashed and dotted lines the estimates of the SSC and IC/CMB
X-ray flux, respectively.}\label{n3sed}
\end{figure}

We have fitted three models to the radio through X-ray flux
density distributions of the seven regions shown in Figures~\ref{s12sed} through~\ref{n3sed}: a
broken power-law synchrotron model with a high energy cut-off,
a synchrotron self-Compton model (SSC) in which the X-ray emission is
the result of inverse-Compton scattering of the synchrotron radio
photons off the relativistic electrons, and a model of inverse
Compton scattering of CMB photons (IC/CMB) off the 
radio synchrotron emitting relativistic electrons.
Other sources of seed photons in the inverse-Compton scenario
including optical photons from the stellar population of
the host galaxy or beamed optical/IR photons from a hidden nuclear jet are implausible
because of the distance of the hot spots from the nucleus
(see \citet{mjh98} for details).
For example, the energy density of IR photons emitted from
the nucleus (based on the nuclear flux measured in the
{\em Spitzer} images) is roughly two orders of magnitude below the energy
density of the CMB at the distance of the hot spots.

The relevant model parameters for the synchrotron, IC/CMB, and SSC curves
shown in Figures~\ref{s12sed} through~\ref{n3sed} are summarized in
Table~\ref{fitptab}.
We assume a power-law electron energy distribution for the synchrotron
fits with $\gamma_{min}$=100 and $\gamma_{max}$=10$^{8}$.  We also
allow for a break in the electron energy distribution where the index
of the power law steepens.  We choose to plot synchrotron models for S3 and
S4 with relatively large breaks for consistency with S1 and S2.  For the
latter two regions, the large change in power-law index
is required by the optical and IR data.  
The flux density distributions of S3 and S4 could be modeled with
a smaller change in index which would be roughly consistent with the optical
and IR limits, although the lack of detections in these bands prevent a
definitive conclusion.  We emphasize that in regions
S3 and S4, we must be averaging over considerable substructure in the
synchrotron scenario because of the short lifetime of the particles.
The broken power-law model thus should be viewed as a parameterization of the
data and does not imply a single-zone model.
For the IC/CMB and SSC fits, we assume equipartition with no contribution 
from relativistic protons.  For all regions other than S1, we assume a cylindrical
geometry in the plane of the sky with the diameter of the cylinder equal to the short 
axis of the region, and the height equal to the long axis.  
The IC/CMB fits roll off at low frequency because of the choice
of $\gamma_{min}$.

The SSC calculation is sensitive to the assumed geometry of the spot,
and high-frequency radio observations of S1 clearly indicate the presence of
compact structures smaller than the box chosen for X-ray analysis.
We consider two models of the geometry of S1 for SSC calculations.
Firstly, we assumed that all the radio emission in the S1 region
comes from a uniform-density sphere of radius $0\farcs5$. This
value was chosen as a reasonably compromise
between the the 15 GHz radio morphology and the much smaller
value of $0\farcs17$ used in the analysis of \citet{mjh98},
who based it on the modeling of \citet{mei89}.
The results of this simple model are plotted in Figure~\ref{s12sed}. As
the hotspot region is clearly resolved in both radio and X-ray, we
next used the code of \citet{mjh02} to consider a more
complex two-zone model which better matches the structure of the
hotspot in the highest-resolution 15-GHz observations. This model
consisted of a section of a thin spherical shell with uniform
electron density embedded in a region bounded by a paraboloid with
a linear electron density gradient: the magnetic field strengths in
the two components were separately normalized using their observed
15-GHz flux densities, and other electron spectral parameters were
as described above. The predicted X-ray flux density of this model
(which is dominated by the more extended filled paraboloid region,
as a thin shell is a poor source of SSC emission) is roughly a
factor 2 below the estimate derived from the spherical geometry. We
conclude that the largest uncertainty in the SSC flux from S1 is due 
to the assumed geometry, but that the observed X-ray flux is
consistent with the SSC prediction with only a small departure from equipartition.

Although the observed X-ray flux from S1 is consistent with the
SSC prediction, the X-ray flux densities
of regions S2, S3, S4, N1, and N2 are one to two orders 
of magnitude larger than predicted in either
the SSC and IC/CMB scenarios if the features are in equipartition.
(see Figures~\ref{s12sed} through~\ref{n3sed}).
The larger region N3 is reasonably well described by the IC/CMB model
with a small departure from equipartition.
The radio through optical flux density distributions of the regions
S2 through S4 indicate a significant steepening of the spectral index in the optical
region, so that the observed X-ray emission cannot be described by an
extrapolation of a simple single-component synchrotron model for
these regions of the SHS; if this extrapolation is
plotted (e.g. in Fig.~\ref{s12sed}) it would clearly violate the optical
constraints on the flux density distributions.
In fact, it has been argued that this radio to optical spectral steepening
strengthens the case for a high-energy cut-off in the electron
energy distribution \citep{mei89,mjh98}.
Thus, while the X-ray emission from S1 is consistent with the SSC mechanism,
neither of the two canonical emission mechanisms provide
a satisfactory explanation for the X-ray emission observed in
the other regions of the SHS of 3C 33.
The X-ray emission from the two compact regions of the NHS (N1 and N2) is orders of
magnitude above the SSC prediction, but is well described by a
simple single-component synchrotron model (shown on Figure~\ref{n12sed}).

There are several important implicit assumptions in the
various IC scenarios that may not be appropriate for the regions of
3C 33 where SSC and IC/CMB are nominally inadequate.
Such failures could account for the discrepancies between the
observed and predicted X-ray flux densities.  In particular, it is possible that
these regions are not in equipartition, that
protons make a significant contribution to the energy density, or that
the filling factor of the relativistic plasma is less than unity.
Only a comparatively small departure from equipartition, a factor of
$\sim$4 in magnetic field (in the sense that $B \approx B_{\rm
eq}/4$), is required for regions S2 through S4 
to be fitted with an IC model.
However, the departure is considerably larger for 
the two compact regions of the NHS ($B \approx B_{\rm eq}/14$).
The equipartition expectations are only reduced if
there is an energetically significant proton population. In principle
arbitrarily large SSC fluxes may be produced by
low filling factors, but in practice these would have to be implausibly
low, with volume filling factor values $\sim 10^{-9}$, to explain the
two orders of magnitude discrepancy between the prediction and the
observations in components such as S4.
These issues are explained in more detail in \citet{mjh04,kat05}.

Relativistic beaming is often invoked to explain the X-ray emission
of jets of radio quasars and the hot spots of FR II radio galaxies in
the various IC scenarios \citep{gk03}.
This is untenable in regions S2-S4 and N1-N2 of
3C 33 for at least five reasons.  First, and most
importantly, X-ray emission has been detected from both the
NHS and SHS.  It is unlikely that both are moving towards us at 
relativistic velocity near our line
of sight.  Second, the X-ray nucleus is heavily obscured,
suggesting that the jets lie close to the plane of the sky.
Third, the optical nucleus is classified as a narrow emission line
radio galaxy, supporting the previous point.  
Fourth, core prominence (the ratio of 5 GHz luminosity of the
core to 178 MHz flux of the entire source)
is often used to estimate the role of relativistic beaming
\citep{orr82, mor97, mjh04}.
The core prominence of the 3C 33 nucleus is low, supporting
the conclusion that beaming is not important.
Fifth, VLBA observations of the core of 3C 33 show only
a small (factor of 2 at a distance of 5 mas from the core with the brighter
component aligned with the SHS) asymmetry
in the ratio of the flux density of the jet/counterjet at 5 GHz \citep{gio05}.
Thus, we conclude that relativistic beaming is not an important factor.

As an alternative, we consider a scenario in which
internal motions in the regions at relativistic velocities
could account for the observed X-ray emission.
In this model, the radio emission originates in two distinct
components, one which follows the general flow of the hot
spots (i.e. non-relativistic in or near the plane of the sky)
and a second component that has been deflected close to our
line of sight at relativistic velocity.  To evaluate the
plausibility of this model, we consider knot S4.  If we
assume that all of the radio emission from this region
originates in the beamed component and it occupies the
entire volume of S4, a bulk Lorentz factor, $\Gamma$, of $\sim$2
or greater is required, and the direction of motion must
be close to the line of sight ($\theta\sim 1/\Gamma$).
Assuming a smaller volume would only push $\Gamma$ higher.
There is some evidence for mildly relativistic flows behind
FR II hot spots, although nothing of this magnitude.  We conclude
that it is unlikely that the X-ray emission from these regions
can be accounted for by relativistic motions of sub-components.

It is also extremely unlikely that the X-ray emission from these regions is due
to hot gas that has been swept up, since the host galaxy of 3C 33
resides in a poor environment.
Assuming that the X-ray emission of the SHS is from a thermal plasma
with $k_BT$=1.5 keV and $Z$=0.5 times Solar, the density of the
gas would be $\sim$10$^{-2}$ cm$^{-3}$ assuming a uniform filling
factor.  The pressure and mass of this gas would be 
$\sim$6.2$\times$10$^{-11}$ dyn cm$^{-2}$ and $\sim$10$^9$ $M_\odot$,
respectively.  
The gas pressure of this putative shell would exceed the equipartition pressure
of the diffuse regions behind the hot spot (S1) by a factor of $\sim$6 \citep{rud88}.
This is not implausibly large as the equipartition pressure of
lobes and jets of FR I radio galaxies is often less than the pressure
in the ambient medium \citep[e.g.  the inner radio lobes of Cen A, see][]{kra03}, although it would be
unusual for an FR II.  If the extended diffuse X-ray emission described in subsection~3.3 is
from an extended corona, the mass of gas in the NHS and SHS would each
be $\sim$5\% of the total gas mass, which is implausibly large.

The only other possibility is that the X-ray emission is synchrotron
radiation from a population of ultra-relativistic electrons, although
the observed flux density distributions cannot be simply 
described with single or continuous injection
models.  There are, in fact, several arguments that support
the multi-zone synchrotron hypothesis for regions S2, S3, S4, N1, and N2.  
First, the radiative lifetime of the ultra-relativistic
particles is a few tens of years in the equipartition magnetic fields
of the regions.  All of the knots are resolved in the X-rays,
so the particles must be re-energized along the length of the regions.  
Thus, in each knot there must be many sites of particle acceleration.
There is therefore no reason to expect that a single injection model
will accurately describe the broad band flux density distribution from such a scenario.
Second, it is clear from Chandra observations of nearby FR I jets, whose
radio through X-ray emission is believed to be synchrotron,
that the morphological relationships between the radio, IR, optical, and
X-ray synchrotron emitting electrons are complicated \citep{kra01,mjh01}.

It is clear, however, that the simple model described in \citet{mjh04}, in
which a single continuous-injection spectrum describes the radio
through X-ray spectrum of these regions, does not apply to 3C 33. The
resolved nature of the X-ray emission, coming from throughout the
bright 10-kpc SHS, in fact makes it clear that a model in which the
X-rays are generated by electrons accelerated at a single location is
simply not appropriate in this object. At the same time, both the 
resolved nature of the X-ray
emission and the fact that 3C 33 lies in the plane of the sky rule out
explanations involving beaming, such as those used by \citet{tav05}
for their quasar hot spots.  Given the known complex
distributed particle acceleration properties of low-power radio jets,
we feel it is reasonable to suggest that the SHS of 3C 33 contains many
discrete regions of particle acceleration and that incompatibility
with a one-zone continuous injection model is not inconsistent with
the X-rays having a synchrotron origin.
In fact, a multiple component synchrotron model for the SHS was required by
\citet{mei97} to explain the radio to optical spectral index and the
optical morphology.  Deep, high spatial resolution observations in the
optical and the near-UV may offer the best prospect of testing this picture.

Therefore, both the synchrotron and SSC processes are important
in the SHS of 3C 33.  The equipartition magnetic field
strength is $\sim$195 $\mu$G for region S1 (and the field strength
would be larger for a smaller assumed volume), and between 25-80 $\mu$G for
regions S2 through S4.  Thus, the rate of synchrotron energy loss
in region S1 is 5-64 times greater than in the rest of the
regions.  There is no X-ray synchrotron emission from S1 because
the magnetic field is too large \citep{mjh04}.  
Extended regions of X-ray emission, where the SSC and IC mechanisms are out
of the question, are strong evidence for a distributed particle
acceleration process like that seen in low-power jets. This suggests
that a simple choice between an SSC or an IC origin for the emission
mechanism may not be appropriate for other, more distant, hot spots
for which we do not have the high linear resolution that we have in
3C 33.

\subsection{Diffuse Emission Co-Incident with Radio Lobes}

We consider two possible models for the origin of the diffuse
X-ray emission coincident with the radio lobes: inverse-Compton scattering of CMB photons off
the relativistic electrons in the diffuse radio features behind the
hot spots and thermal emission from hot gas.
The quality of the data is not good enough for us to differentiate
with certainty between these possibilities.

\subsubsection{IC/CMB}

X-ray emission from the extended lobes of radio galaxies is
generally attributed to inverse-Compton scattering of cosmic microwave
background (CMB) photons by the low-energy electrons in the lobe \citep{fei95,kan95}.
Additionally, it has been shown for large samples of FR II radio sources that
the X-ray fluxes measured from the lobes generally lie within a factor
of a few of the prediction from the inverse-Compton process assuming
equipartition magnetic fields \citep{kat05,cros05}.  We calculate the expected
inverse-Compton fluxes from the lobes on the assumption of
equipartition using the code of \cite{mjh98} and with the same
assumptions as \cite{cros05} (lobes are treated as uniform cylinders,
$\gamma_{\rm min} = 10$, $\delta=2$, and a break in the electron
spectrum is applied to match the radio data), taking normalizing flux
densities for the lobe regions, excluding the NHS and SHS, from
unpublished archival low-frequency VLA (1.4 GHz, observation
AL0146) and GMRT (610 and 244 MHz, observation
number 2070) data.  3C 33 is unusual in that the flux
density even at low frequencies is dominated by the bright hot spots, not
the lobes.  This calculation predicts flux densities of 0.7 nJy for each lobe at
equipartition, a factor 1.9 and 2.6 below the observed values. As
\cite{cros05} find that this factor is typically $\sim 2$ for the
sources they study, 3C\,33's lobe X-ray emission gives results
consistent with those seen in other FR IIs.  The X-ray morphology does
not precisely match the radio morphology, although the low X-ray surface
brightness and heavy smoothing of the data prevent a definitive statement.
Thus it is at least plausible that this extended X-ray emission is due to
inverse-Compton emission.

\subsubsection{Thermal}

The other possibility is that this X-ray emission
is from an extended hot gas corona.  We consider two models for
the distribution of the gas, a uniform density sphere and a shell
surrounding the radio lobes.  Modeling the emission region as
a uniform-density sphere of radius 1$'$ (68.4 kpc) with the best-fit temperature
of the thermal model, we find a gas density of 7$\times$10$^{-4}$ cm$^{-3}$ 
and a total mass of 2$\times$10$^{10}$ $M_\odot$. 
A more realistic density profile (i.e. a beta-model) would only change
these numbers by a factor of order unity.
This is a relatively small amount of gas, probably less than the mass
of stars in the host galaxy.  The optical luminosity of the host
galaxy ($M_B$=-20.7) implies a stellar mass of $\sim$9$\times$10$^{10}$ $M_\odot$
assuming a mass to light ratio of 6 \citep{bin87}.
The total thermal energy and pressure of this gas is $\sim$1.3$\times$10$^{59}$ ergs
and $\sim$3.2$\times$10$^{-12}$ dyn cm$^{-2}$, respectively.
The thermal energy of the gas is not particularly large compared with
the mechanical energy of many powerful radio galaxies such as 3C 33 \citep{bir04,kra06}.
Any atmosphere of this mass, or less, is likely to have been shock heated by the radio outflow,
and so would no longer be bound to the relatively shallow gravitational potential of the host
galaxy of 3C 33.

Alternatively, the X-ray emission coincident with the lobes
could be from a shell of gas that has been swept up and compressed by the
inflation of the radio source. Modeling the northern emission region as a spherical
shell with a radius of 62 kpc and thickness of 10 kpc, with the
best-fitting temperature of the thermal model, we find a proton
density in this shell of $6.0 \times 10^{-3}$ cm$^{-3}$, and a total mass of $7
\times 10^{8}$ $M_\odot$.  For a thinner, 1-kpc thick shell, the mass
is $2 \times 10^{8}$ (similar numbers are obtained for the southern
lobe). As a comparison, we estimated the amount of mass expected to
have been swept up by the expanding radio lobe, assuming the $\beta$
model parameters and temperature of Section 3.4, and assuming an
expanding cone of opening angle 48 degrees and an outer radius of 90
kpc, which gave a mass of $6 \times 10^{8}$ M$_{\sun}$. It is
therefore possible that this emission originates from compressed gas
that has been swept up by the expanding radio source; this gas may
also have been shock heated, since the radio lobes are overpressured
(see Section 4.3) and therefore expanding supersonically at least in
their outer parts. 

We note however that if the lobe-related emission is
interpreted as entirely thermal in origin, then the absence of
inverse-Compton emission from the radio lobes would require them to be
magnetically dominated, which is not the case for the general
population of FR II radio sources \citep{cros05}. We therefore argue that the majority
of the lobe-related X-ray emission is likely to be inverse-Compton in
origin; however, we cannot rule out some contribution from thermal
emission related to radio-source environmental impact.

\subsection{Large Scale Environment and Radio Lobe Dynamics}

3C\,33 appears to have a poor environment with a steep density and 
pressure gradient in the regions surrounding the radio lobes. 
We compared the internal minimum pressure of the
radio lobes with the external pressure from this environment in order
to investigate the dynamics of radio-lobe expansion. The radio lobes
were modeled assuming equipartition and the same electron spectrum
and normalizing flux densities as in Section 4.2.1, and found to have
minimum internal pressures of $9 \times 10^{-13}$ dyn cm$^{-2}$ (this value is
slightly higher if the electron density implied by an inverse-Compton
interpretation for the lobe X-ray flux is used: see Section 4.2.1). 
Our estimate of the external gas pressure depends on whether the observed
diffuse X-ray emission on large scales is thermal or IC/CMB.
If we assume the emission is thermal, the X-ray data beyond 20$''$ from the nucleus
are consistent with a flat beta-model profile of low surface brightness.
For example, a component with $\beta$=0.5 and $r_{c}$=20 kpc,
normalized to the observed {\it Chandra} count density at
50 arcsec, would have external pressures at the midpoints of the radio
lobes of $\sim$2$\times$10$^{-12}$ dyne cm$^{-2}$, consistent with
approximate pressure balance as seen in other sources for which similar
analyses have been performed (e.g. \citealt{mjh00,mjh02,cro04,bel04} among many others).
On the other hand, if we simply extrapolate the best fit beta-model
parameters derived for the galactic scale emission (second and third
paragraphs of section 3.3), the lobes are greatly overpressured relative
to the ambient medium.  At the approximate midpoint
of the northern lobe, a distance of $\sim 69$ kpc, the external
pressure is $3.7^{+6.4}_{-0.6} \times 10^{-14}$ dyn cm$^{-2}$ (with the
uncertainty dominated by uncertainty in the $\beta$-model parameters);
at the (projected) midpoint of the southern lobe, $\sim 52$ kpc from
the nucleus, the external pressure was found to be $7.9^{+12.0}_{-1.2}
\times 10^{-14}$ dyn cm$^{-2}$. Although both lobes are likely to be in pressure
balance at their inner edges, with this model for the environment,
which has an exceptionally steep density gradient, the northern lobe appears to be
overpressured by at least an order of magnitude halfway along its
length, and both lobes must be overpressured at their outer edges. 

The teardrop shape of the SHS has often been interpreted as shaped by a bow
shock \citep{rud88}, and supports the hypothesis that it is still moving
supersonically (relative to the external medium) through the extended hot corona.
If this interpretation is correct, the opening half-angle
of the shock, $\mu$ is related to the Mach number of the advance
of the jet head by sin($\mu$)$\sim$M$^{-1}$.
The opening angle of the SHS is $\sim$50$^{\circ}$, implying
a Mach number of $\sim$2.4 for the advance speed of the jet head relative
to the ambient medium.
The equipartition pressure of the material just behind the radio hot
spots of the SHS is $\sim$ 10$^{-11}$ dyn cm$^{-2}$ \citep{rud88}.
If the bow shock interpretation is correct, this material should be in rough
pressure equilibrium with the shocked IGM.  The Mach number of the advance
of the jet implies a pressure jump of a factor of $\sim$7 (for $\gamma$=5/3) across
the bow shock in the IGM based on the Rankine-Hugoniot relations \citep{lan89}.  
The ratio of the pressures of the material behind the hot spot to
the unshocked ambient medium is roughly consistent with this value.
Assuming a gas temperature of 1 keV (consistent with the
spectral fit above and the poor environment if all the X-ray emission
is from thermal gas), the velocity (advance speed) of
the jet head is $\sim$1000 km s$^{-1}$.
The radio outburst therefore commenced $\sim$7$\times$10$^{7}$ years ago if
the jet has not slowed appreciably.
Conservatively estimating that the mechanical power deposited into the
gas equals the X-ray luminosity of the nucleus, the jets have
put $\sim$2$\times$10$^{59}$ ergs into the gas, a number roughly
equal to the thermal energy of the gas (if all the diffuse X-ray emission
has a thermal origin).  We emphasize that this is a very
conservative estimate:  if in fact the mechanical
power of the jets/lobes is 10-100 times greater than the nuclear luminosity as
has been observed in several radio galaxies in denser environments \citep{bir04},
the energy deposited in the gas could be considerably larger.
The powerful radio outburst has at least doubled the energy density of any
gas within $\sim$70 kpc of
3C 33.  It (the corona) is now likely unbound and in the process of
being blown out of the galaxy.  The diffuse X-ray emission may be the remnant 
of the group gas that has escaped/is escaping as a wind.

\section{Conclusions}

We have detected X-ray emission from both the north and south hot spots
of the radio galaxy 3C 33.
The X-ray emission from the tip of the SHS (region S1) is consistent with the SSC process.
For the larger regions behind S1 and for the two compact regions
of the NHS (N1 and N2), the observed X-ray flux
is an order of magnitude or more above the SSC prediction.
All inverse-Compton scattering models
are rejected unless the hot spots are far from equipartition, and 
relativistic beaming cannot be significant as X-ray emission is detected
from both hot spots and the nucleus is heavily absorbed.
The radio through optical flux density distributions show a gradual spectral steepening, 
indicating that the observed X-ray flux from regions S2-S4 cannot be explained 
by a simple power-law extrapolation of the longer wavelength spectrum.  
We conclude that the X-ray emission is most likely synchrotron, although
the measured flux densities lie above a simple extrapolation of the
radio/IR/optical flux densities, or predictions from single or
continuous injection models.
The distributed nature of the X-ray emission throughout the diffuse radio
features behind the compact radio hot spot, not
just coincident with the radio peaks as seen in, for example,
Cygnus A \citep{wil00}, is analogous to the synchrotron emission
seen in FR I jets.

The hot spots of 3C 33 thus provide strong evidence for the presence
of synchrotron and SSC components within the same hot spots, and for multiple synchrotron
components in FR II hot spots where the effects of relativistic beaming are unimportant.
If this radio galaxy were observed at lower spatial resolution and the
entire volume of the SHS used for the SSC calculation, we would conclude
that the observed X-ray emission from the SHS far exceeded the SSC prediction
for the feature as a whole.  In fact, both emission mechanisms are important,
and SSC is the dominant mechanism in the most compact feature.  This should warn
us about stretching assumptions about uniform filling factors for the hot
spots of more distant radio galaxies too far.  

High resolution UV images of the two hot spots could be made to further
evaluate the conclusions of this paper.
Similarly detailed, multiwavelength (IR and optical) studies of the NHS should
also be made to confirm the synchrotron origin.
Detailed optical studies of the NHS have not been previously
undertaken probably because of the large difference in radio
flux density relative to the SHS.
Higher spatial resolution optical (i.e. {\em HST}) observations of the NHS may be
particularly useful in constraining the morphology of this emission.

It is now clear that the SSC explanation for the X-ray emission from
the radio hot spots of FR IIs is not the complete picture, and that synchrotron emission
must be important in at least some of these sources.  We have
demonstrated that at least in the case of 3C 33, relativistic beaming
is unlikely to play an important role.  The role of beaming in general,
however, is not known, and can only be addressed by studying statistically complete
samples.  The relative importance of synchrotron versus SSC X-ray
emission in FR II hot spots could be conclusively resolved if a complete sample 
of such radio galaxies were observed by {\em Chandra}.
We believe it is crucial to {\em Chandra}'s legacy that such a sample
is observed.

\acknowledgements

We would like to thank the 3CRR low-z consortium for providing us
their {\em Spitzer} data of 3C 33 prior to publication.
We would also like to thank the anonymous referee for detailed
comments that improved this paper.
This work was supported by Chandra contracts NAS8-38248, NAS8-39073,
the Chandra X-ray Center, the Smithsonian Institution, and the
Royal Society.  The National Radio Astronomy Observatory is a facility 
of the National Science Foundation operated under cooperative agreement 
by Associated Universities, Inc.

\clearpage

\clearpage

\clearpage

\clearpage

\clearpage

\clearpage 

\clearpage 

\clearpage

\clearpage

\clearpage

\clearpage

\clearpage

\clearpage

\clearpage

\clearpage 

\begin{table}
\begin{tabular}{llllll}
\hline
Observation ID & Hotspot & Frequency & Configuration & Date & Map resolution\\
&&(GHz)&&&(arcsec)\\
\hline
AR130&S&4.9&A&1986 Jun 05&$0.43 \times 0.38$\\
AR148&N&4.9&B&1986 Jul 13&$1.37 \times 1.29$\\
AR148&N&15.0&B&1986 Jul 13&$0.45 \times 0.40$\\
AR148&S&4.9&B&1986 Jul 13&$1.37 \times 1.29$\\
AR148&S&15.0&B&1986 Jul 13&$0.45 \times 0.40$\\
AR223&S&15.0&B&1990 Sep 14&$0.45 \times 0.40$\\
AR208&S&15.0&A&1990 Apr 24&$0.19 \times 0.12$\\
\hline
\end{tabular}
\caption{Details of VLA data sets retrieved from the public archive.
  The resolutions quoted are the FWHM major and minor axes of the
  restoring elliptical Gaussians.}
\label{vla}
\end{table}

\clearpage

\begin{table}
\begin{tabular}{|c|c|c|c|c|c|c|}\hline
Region & RA & DEC & Dimensions & Angle & Rate & Flux Density \\ \hline\hline
N1  & 01:08:56.4 & +13:22:22.9 & 2.5$''$$\times$2.5$''$ & 50$^\circ$ & 2.7$\pm$0.8 & 0.27$\pm$0.08 \\ \hline
N2  & 01:08:56.3 & +13:22:24.9 & 2.5$''$$\times$2.5$''$ & 50$^\circ$ & 2.0$\pm$0.7 & 0.19$\pm$0.07 \\ \hline
N3  & 01:08:55.8 & +13:22:13.7 & 12.4$''$$\times$24.2$''$   &  45$^\circ$ & 2.5$\pm$0.7 & 0.24$\pm$0.07 \\ \hline
DN1 & 01:08:55.4 & +13:21:13.0 & 1.4$'$$\times$1.1$'$   & 85$^\circ$ & 11.8$\pm$1.0 & 1.3$\pm$0.1 \\ \hline
S1  & 01:08:50.4 & +13:18:26.3 & 2.7$''$$\times$2.5$''$ & 40$^\circ$ & 1.4$\pm$0.6 & 0.14$\pm$0.06 \\ \hline
S2  & 01:08:50.5 & +13:18:28.5 & 3.0$''$$\times$3.0$''$ & 40$^\circ$ & 3.1$\pm$0.9 & 0.32$\pm$0.09 \\ \hline
S3  & 01:08:50.4 & +13:18:33.0 & 4.7$''$$\times$8.6$''$ & 40$^\circ$ & 3.5$\pm$1.0 & 0.35$\pm$0.10 \\ \hline
S4  & 01:08:50.7 & +13:18:42.4 & 5.9$''$$\times$5.9$''$ & 40$^\circ$ & 4.5$\pm$1.2 & 0.45$\pm$0.12 \\ \hline
DS1 & 01:08:52.8 & +13:19:24.0 & 1.4$'$$\times$1.2$'$   & 36$^\circ$ & 16.1$\pm$1.0 & 1.8$\pm$0.1 \\ \hline
\end{tabular}
\caption{Summary of regions used for flux density distributions.  The dimensions are in arcseconds or
arcminutes (for DN1 and DS1), the angle is the rotation of the box east of north, rate is the net (i.e.
background subtracted) X-ray count rate
in units of 10$^{-4}$ cts s$^{-1}$ in the 0.5-5.0 keV band, and the flux density is in
nJy at 1 keV assuming a power law spectrum with photon index 2 (1.7 for DN1 and DS1) and Galactic 
absorption.  All error bars are 1$\sigma$ counting statistics uncertainties.}\label{regstab}
\end{table}

\clearpage

\begin{table}
{\footnotesize
\begin{tabular}{|l|c|c|c|c|c|c|c|c|c|}\hline
 & \multicolumn{4}{|c|}{VLA (mJy)} & \multicolumn{5}{|c|}{{\em Spitzer} ($\mu$Jy)} \\ \hline
 & 1.477 GHz & 4.9 GHz & 15.0 GHz & 231 GHz & 23.9 $\mu$m & 7.87 $\mu$m & 5.73 $\mu$m & 4.49 $\mu$m & 3.55 $\mu$m \\ \hline\hline 
S1 &       & 622. & 270. & 34. & 1554$\pm$119 & 320$\pm$19 & 207$\pm$21 & 146$\pm$9 & 99$\pm$6 \\ \hline
S2 &       & 506. & 169. &     &  777$\pm$90  & 183$\pm$21 & 123$\pm$20 &  67$\pm$6 & 47$\pm$4 \\ \hline
S3 & 1670. & 617. & 158. &     &  & $<$82 & $<$68 & $<$20 & $<$11 \\ \hline
S4 & 435.  & 140. &      &     &  & $<$45 & $<$77 &  $<$9 &  $<$8 \\ \hline
N1 &       & 64.  &  24. &     &  196$\pm$137 &  22$\pm$20 &  20$\pm$18 &  18$\pm$5 & 12$\pm$4 \\ \hline 
N2 &       & 60.  &  22. &     &  178$\pm$125 &  26$\pm$22 &  32$\pm$19 &   9$\pm$5 &  6$\pm$4 \\ \hline
N3 & 1360. & 460. &      &     &  & $<$66 & $<$115 & $<$13 & $<$12 \\ \hline
\end{tabular}
}
\caption{Summary of radio and {\em Spitzer}/IR flux density measurements.  The
uncertainties on the {\em Spitzer} points are 1$\sigma$.  The upper limits on
the {\em Spitzer} IRAC data are 3$\sigma$ (statistical uncertainties only).}\label{fdsumtab}
\end{table}

\clearpage

\begin{table}
{\footnotesize
\begin{tabular}{|l|c|c|c|c|c|c|c|c|}\hline
 & \multicolumn{5}{|c|}{Optical ($\mu$Jy)} & \multicolumn{2}{|c|}{Galex ($\mu$Jy)} & \multicolumn{1}{|c|}{Chandra (nJy)} \\ \hline
 & 6696 \AA & 6500. \AA & 6493.5 \AA & 4800 \AA & 4497.8 \AA & 2297\AA & 1524 \AA & 1 keV \\ \hline\hline
S1 & 4.5 & 3.9$\pm$2.8 & 3.7 & 1.3 & 3.4$\pm$0.7 & $<$2.0 & $<$6.1 & 0.14$\pm$0.06 \\ \hline
\end{tabular}
}
\caption{Summary of optical, UV, and X-ray flux density measurements for S1.  Optical points are
taken from \citet{sim86,cra87,mei89}.  
Uncertainties on the X-ray data points are 1$\sigma$ counting statistics only.  
The {\em GALEX} points are not detections, only the 3$\sigma$ upper limits (see text for complete
discussion).}\label{fdsumtaboptx}
\end{table}

\clearpage

\begin{table}
\begin{center}
\begin{tabular}{|c|c|c|}\hline
 & Model 1 & Model 2 \\ \hline
 & Power-Law plus Reflection Model & Two Power-Law Model \\ \hline
\multicolumn{3}{|c|}{Primary Power law} \\ \hline
Photon Index &  1.7$\pm$0.2 & 1.7$\pm$0.2 \\ \hline
$N_H$        &  (5.4$\pm$0.8)$\times$10$^{23}$ cm$^{-2}$ & (3.5$\pm$0.4)$\times$10$^{23}$ cm$^{-2}$ \\ \hline
Flux         &  2.4$\times$10$^{-11}$ ergs cm$^{-2}$ s$^{-1}$ & 2.9$\times$10$^{-11}$ ergs cm$^{-2}$ s$^{-1}$ \\ \hline
Luminosity   &  2.0$\times$10$^{44}$ ergs s$^{-1}$ & 2.5$\times$10$^{44}$ ergs s$^{-1}$ \\ \hline
\multicolumn{3}{|c|}{Fe K$_\alpha$ line} \\ \hline
Centroid     & 6.38$\pm$0.04 keV & 6.37$\pm$0.04 keV \\ \hline
Flux         & 2.4$\times$10$^{-5}$ photons cm$^{-2}$ s$^{-1}$ & 3.1$\times$10$^{-5}$ photons cm$^{-2}$ s$^{-1}$ \\ \hline
EW           & 210$\pm$70 eV & 170$\pm$65 eV \\ \hline
\multicolumn{3}{|c|}{Thermal Component (APEC)} \\ \hline
$k_BT$       & 0.42$\pm$0.06 keV & 0.46$\pm$0.05 keV \\ \hline
$Z_{Mg}$     & 2.0 & 1.0 \\ \hline
$Z_{Si}$     & 2.0 & 1.0 \\ \hline
$Z_{Fe}$     & 0.15$\pm$0.05 & 1.0 \\ \hline
Flux         & 5.9$\times$10$^{-14}$ ergs cm$^{-2}$ s$^{-1}$ & 5.6$\times$10$^{-14}$ ergs cm$^{-2}$ s$^{-1}$  \\ \hline
Luminosity   & 5.1$\times$10$^{41}$ ergs s$^{-1}$ & 4.8$\times$10$^{41}$ ergs s$^{-1}$ \\ \hline
 & Reflection Component & Second Power Law \\ \hline
Photon Index &  1.7 (fixed) & 2.0 (fixed) \\ \hline
Relative Norm. & 2.6 & \\ \hline
$N_H$        &   &  1.2$\pm$0.4$\times$10$^{22}$ cm$^{-2}$ \\ \hline
Flux         &  1.6$\times$10$^{-12}$ ergs cm$^{-2}$ s$^{-1}$ & 8.3$\times$10$^{-13}$ ergs cm$^{-2}$ s$^{-1}$ \\ \hline
Luminosity   &  1.2$\times$10$^{43}$ ergs s$^{-1}$ & 7.0$\times$10$^{42}$ ergs s$^{-1}$ \\ \hline
$\chi^2$/dof & 27.3/18 & 48.1/19 \\ \hline
\end{tabular}
\end{center}
\caption{Summary of best fit spectral parameters of the nucleus of 3C 33.  All
fluxes are unabsorbed in the 0.1-10.0 keV band, all luminosities are
in the 0.1-10.0 keV band.  The equivalent width (EW) of the Fe K$_\alpha$
line and the normalization of the reflection component are
relative to the primary power law.  All uncertainties are at 90\%
confidence for one parameter of interest.}\label{spectab}
\end{table}

\clearpage

\begin{table}
\begin{center}
\begin{tabular}{|c|c|c|c|c|c|c|}\hline
       &  \multicolumn{3}{|c|}{Synchrotron} & \multicolumn{3}{|c|}{SSC} \\ \hline
Region & $P$ & $\Delta$P & $\gamma_{break}$ & Radius & Length & $B_{eq}$ ($\mu$G) \\ \hline
S1     & 2.5 & 2.7 & 2$\times$10$^{5}$ & \multicolumn{2}{|c|}{Section 4.1} & 195 \\ \hline
S2     & 2.7 & 3.2 & 4$\times$10$^{5}$ & 1.5$''$ & 2.96$''$ & 79 \\ \hline
S3     & 2.8 & 2.8 & 2$\times$10$^{5}$ & 2.35$''$ & 8.6$''$ & 53 \\ \hline
S4     & 2.9 & 2.8 & 2$\times$10$^{5}$ & 2.95$''$ & 10.1$''$ & 34 \\ \hline
N1     & 2.4 & 1.3 & 2$\times$10$^{5}$ & 1.25$''$ & 2.5$''$ & 43 \\ \hline
N2     & 2.4 & 1.3 & 2$\times$10$^{5}$ & 1.25$''$ & 2.5$''$ & 42 \\ \hline
N3     & 2.9 & 1.3 & 2$\times$10$^{5}$ & 6.2$''$ & 24.2$''$ & 25 \\ \hline
\end{tabular}
\end{center}
\caption{Summary of model paramters for synchrotron and SSC fits.  Synchrotron
fits assume an electron energy distribution of the form $E^{-P}$ to the break
energy, and a form $E^{-(P+\Delta P)}$ above the break.  The index of the electron
energy distribution, $P$, is determined from the observed spectral index of the
radio flux density, $\alpha$ and is given by $P=2\alpha +1$.
The equipartition magnetic field, $B_{eq}$ is computed assuming no relativistic protons
and unity filling factor.  See section 4.1 for more detailed discussion
of model fits.}\label{fitptab}
\end{table}

\end{document}